\documentclass[aps,%singlecolumn,
showpacs,tightenlines,fleqn]{revtex4}
\usepackage{graphics}

\input epsf %%% eps figures

\begin{document}
\bibliographystyle{apsrev}

\newcommand{\half}{\frac{1}{2}}
\newcommand{\D}{\mbox{D}}
\newcommand{\curl}{\mbox{curl}\,}
\newcommand{\ep}{\varepsilon}
\newcommand{\lleq}{\lower0.9ex\hbox{ $\buildrel < \over \sim$} ~}
\newcommand{\ggeq}{\lower0.9ex\hbox{ $\buildrel > \over \sim$} ~}
\newcommand{\tr}{{\rm tr}\, }

\newcommand{\be}{\begin{equation}}
\newcommand{\ee}{\end{equation}}
\newcommand{\bea}{\begin{eqnarray}}
\newcommand{\eea}{\end{eqnarray}}
\newcommand{\beaa}{\begin{eqnarray*}}
\newcommand{\eeaa}{\end{eqnarray*}}
\newcommand{\Lhat}{\widehat{\mathcal{L}}}
\newcommand{\nn}{\nonumber \\}
\newcommand{\e}{{\rm e}}

\tolerance=5000

\title{String-inspired Gauss-Bonnet gravity reconstructed from
the universe expansion history and yielding the transition from
matter dominance to dark energy}

\author{Guido Cognola$\,^{(a)}$\footnote{cognola@science.unitn.it},
Emilio Elizalde$\,^{(b)}$\footnote{elizalde@ieec.uab.es, also on
leave at Department of Physics \& Astronomy, Dartmouth College, 6127
Wilder Laboratory, Hanover, NH 03755, USA},
Shin'ichi Nojiri$\,^{(c)}$\footnote{nojiri@phys.nagoya-u.ac.jp},\\
Sergei D.~Odintsov$\,^{(b,d)}$\footnote{odintsov@ieec.uab.es, also
at TSPU, Tomsk} and Sergio Zerbini$\,^{(a)}$\footnote{zerbini@science.unitn.it} } \affiliation{
%\address{
$^{(a)}$\,Dipartimento di Fisica, Universit\`a di Trento \\
and Istituto Nazionale di Fisica Nucleare \\
Gruppo Collegato di Trento, Italia\\
\mbox{} \\
$^{(b)}$\,Consejo Superior de Investigaciones Cient\'{\i}ficas
(ICE/CSIC) \\ and Institut d'Estudis Espacials de Catalunya
(IEEC) \\
Campus UAB, Facultat de Ci\`encies, Torre C5-Par-2a pl \\
E-08193 Bellaterra (Barcelona) Spain\\
\mbox{} \\
$^{(c)}$\,Department of Physics \\ Nagoya University, Nagoya 464-8602, Japan \\
\mbox{} \\
$^{(d)}$\,Instituci\`{o} Catalana de Recerca i Estudis Avan\c{c}ats (ICREA) \\
and Institut de Ciencies de l'Espai (IEEC-CSIC) \\ Campus UAB,
Facultat de Ci\`encies, Torre C5-Par-2a pl \\ E-08193 Bellaterra
(Barcelona), Spain\\
}

\begin{abstract}

We consider scalar-Gauss-Bonnet and modified Gauss-Bonnet gravities
and reconstruct these theories from the universe expansion history.
In particular, we are able to construct versions of those theories
(with and without ordinary matter), in which the matter dominated
era makes a transition to the cosmic acceleration epoch. In several
of the cases under consideration, matter
dominance and the deceleration-acceleration transition occur in the
presence of matter only. The late-time acceleration epoch is
described asymptotically by de Sitter space but may also correspond
to an exact $\Lambda$CDM cosmology, having in both cases an
effective equation of state parameter $w$ close to $-1$. The
one-loop effective action of modified Gauss-Bonnet gravity on the de
Sitter background is evaluated and it is used to derive stability
criteria for the ensuing de Sitter universe.

\end{abstract}


\maketitle

\section{Introduction}

Modified gravity is a promising theory that has become a very
attractive gravitational alternative for dark energy (for a recent
review, see \cite{padmanabhan,CST,rev3}. It is, to start, a powerful
scheme. Indeed, depending on the specific model considered, modified
gravity is able to realize any of the proposed scenarios that have
been delimited by the observational constrains leading to cosmic
acceleration: effective phantom models, cosmological constant
theories or quintessence. Also, it can easily account for the
different epochs in the evolution of the unverse, that start to
emerge clearly from the observational data (for a recent review of
these data and their comparison with dark energy models, see
\cite{bagla}). The qualitative understanding of gravitational dark
energy (see \cite{rev3} for a review) is quite simple: some
gravitational terms different from the usual General Relativity ones
may dominate at the very early or very late universe epochs. Thus,
General Relativity seems to be only approximately valid, both at
very early as well as at very late times. To have the possibility to
explain ---in a unified way as modified gravity effects---
fundamental cosmological phenomena such as early time inflation and
late time acceleration, is very appealing. Moreover, modified
gravity has the possibility to solve the coincidence problem, too,
and can also clarify the role of dark matter in the formation and
evolution of the universe.

Among the different classes of modified gravities, a very
interesting one is the family of the string-inspired gravities.
String-inspired scalar-Gauss-Bonnet gravity has been suggested in
Ref.~ \cite{sasaki} as a possibility for gravitational dark energy.
The idea that these theories may lie close to effective models which
may come from fundamental string theories is in itself very
appealing. Some time ago, Gauss-Bonnet (GB) gravity was applied to
the possible resolution of the initial singularity problem
\cite{ART}. Another version of string-inspired gravity, namely the
modified GB or $F(G)$ theory \cite{fGB} can also play the role of
gravitational dark energy. The investigation of different regimes of
cosmic acceleration in such string-inspired gravity models has been
carried out in
refs.\cite{sasaki,fGB,Sami,Mota,Calcagni,Neupane,GB,sami,Cognola:2006eg}.

A very strong theoretical restriction on modified gravity is caused
by the natural emergence of the known classical universe expansion
history. In other words, as in the case of the usual $\Lambda$CDM
cosmology, it must faithfully reproduce, to start with, the sequence
of the well-established cosmological phases: radiation/matter
dominance, deceleration-acceleration transition, and the cosmic
speed-up. Recently, a reconstruction scheme for the case of $f(R)$
gravity has been developed \cite{salvatore,reconstruction,ERE} where
such a cosmological sequence is seen to occur quite naturally for
some models (for a review on reconstruction from the universe
expansion history, see \cite{ERE,varun}). Having in mind the
fundamental importance of the correct description of the past and
the current universes, in this work we develop a reconstruction
program for string-inspired gravity from the universe expansion
history. Using the method developed in refs.\cite{e,sami}, it will
be here demonstrated that scalar-Gauss-Bonnet or $F(G)$ gravity can
be reconstructed for any given FRW cosmology. Moreover, the role of
string-inspired gravitational terms may be quite important, even in
the matter dominated epoch. Examples of exact and/or approximate
$\Lambda$CDM cosmology in the above modified gravity are discussed
too. The stability of such form of FRW cosmology, which naturally
develops an asymptotically de Sitter future is investigated. As de
Sitter universe naturally occurs at the early or late times in such
models, special attention is paid to the de Sitter universe. Using
the results of the calculation of the one-loop effective action of
$F(G)$ gravity in the de Sitter space, the semi-classical stability
of de Sitter space is investigated.

The paper is organized as follows. In the next section we develop
the reconstruction scenario for string-inspired, scalar-Gauss-Bonnet
gravity. Elaborating on the approach of Refs.~\cite{e,sami}, it is
shown that the cosmological sequence of matter dominance,
deceleration-acceleration transition and cosmic acceleration may
occur in the specific version of that theory with some given
potentials. It is also shown that an exact $\Lambda$CDM cosmology
can indeed be reconstructed from such theory. Section three is
devoted to the reconstruction of $F(G)$ gravity from the universe
expansion history. For the same two classes of FRW universes (an
approximate and an exact one), $\Lambda$CDM cosmology is derived in
versions of the theory where this cosmology naturally occurs. By
introducing perturbations around such cosmological solutions, the
stability of those two FRW universes is investigated at the
classical level. It is shown that the very early universe may be
unstable, which opens the very interesting possibility of a natural
exit from inflation, while the late stage, e.g. the asymptotically
de Sitter universe, can be stable. Sect.~4 is devoted to the study
of the same reconstruction scenario for $F(G)$ gravity in the
presence of several types of usual matter. Two specific versions of
$F(G)$ gravity with matter are constructed where the matter
dominance and deceleration-acceleration transition occur only {\it
with matter}. After that, the accelerating universe becomes
asymptotically (or even exactly) the de Sitter one.

In Sect.~5 we discuss the following problem: does it exist any
scenario which may still improve the above situation, with the
emergence of the matter dominated era for scalar-Gauss-Bonnet or
$F(G)$ gravity? It is shown that such scenario may be realized: by
adding the compensating dark energy which is relevant in the matter
dominance epoch mainly, the cosmological sequence of matter
dominance, deceleration-acceleration transition, and cosmic
acceleration, do occur in the versions of such theories which,
otherwise, do not contain the matter dominance. Some discussion and
outlook are presented in the last section.

Appendix A is devoted to the calculation of the one-loop effective
action of $F(G)$ gravity on a de Sitter background. Some technical
remarks about the appearance of the multiplicative anomaly in those
calculations are done. In Appendix B, this one-loop effective action is
applied to study the semi-classical stability of de Sitter space in
$F(G)$ gravity. This is an alternative method to investigate
stability, which is advantageous as compared with the very involved
one that uses cosmological perturbations. Numerical calculations
provide examples of modified Gauss-Bonnet gravity which have a
stable de Sitter vacuum solution.

\section{Reconstruction of the scalar-Gauss-Bonnet theory from the
universe expansion history}

In the present section, we will show how the string-inspired,
scalar-Gauss-Bonnet gravity theory, proposed as a dark energy model in
Ref.~\cite{sasaki}, can indeed be
reconstructed, for any requested cosmology. The starting action is
\be
\label{GBany1}
S=\int d^4 x \sqrt{-g}\left[ \frac{R}{2\kappa^2} - \frac{1}{2}\partial_\mu \phi
\partial^\mu \phi - V(\phi) - \xi_1(\phi) G \right]\ .
\ee
Here $G$ is the Gauss-Bonnet invariant and the scalar field $\phi$ is canonical in (\ref{GBany1}).
We now assume that the FRW universe with scale factor $a(t)$:
$ds^2=-dt^2 + a(t)^2\sum_{i=1,\cdots 4}\left(dx^i\right)^2$
and the scalar field $\phi$ only depend on $t$ : $\phi=\phi(t)$. 
%%%%%%%%%%%%%%%
These assumptions are based on the observational data indicating that the spatial part of the universe is
flat and the universe could be regarded to be homogeneous on large scales. 
In the early universe, the small inhomogeneity could have played an important role. For example, the large scale 
structure of the present universe  could be generated from the  very small initial fluctuation 
and the gravitational instability for the perturbation. 
In this paper, however, we are interested in the behavior of the universe on the scales larger than the `large' 
scale structure and we now assume the homogeneity of the universe.
%%%%%%%%%%%%
Hence, we now drop the spatial coordinate ($x^i$) dependence of $a$ and $\phi$.
In the following, the Hubble rate $H$ is defined by $\dot a/a$.

Then the FRW equations look like \cite{sasaki,sami}:
\bea
\label{GBany4}
0&=& - \frac{3}{\kappa^2}H^2 + \frac{1}{2}{\dot\phi}^2 + V(\phi) + 24 H^3
\frac{d \xi_1(\phi(t))}{dt}\ ,\\
\label{GBany5}
0&=& \frac{1}{\kappa^2}\left(2\dot H + 3 H^2 \right) + \frac{1}{2}{\dot\phi}^2 - V(\phi)
- 8H^2 \frac{d^2 \xi_1(\phi(t))}{dt^2} - 16H \dot H
\frac{d\xi_1(\phi(t))}{dt} - 16 H^3 \frac{d \xi_1(\phi(t))}{dt}\ .
\eea
and the scalar field equation is
\be
\label{GBany6}
0=\ddot \phi + 3H\dot \phi + V'(\phi) + \xi_1'(\phi) G\ .
\ee
Now $G=24\left(\dot H H^2 + H^4\right)$, and
combining (\ref{GBany4}) and (\ref{GBany5}), one gets
\be
\label{GBany7}
0=\frac{2}{\kappa^2}\dot H + {\dot\phi}^2 - 8H^2 \frac{d^2
\xi_1(\phi(t))}{dt^2} - 16 H\dot H \frac{d\xi_1(\phi(t))}{dt} + 8H^3 \frac{d\xi_1(\phi(t))}{dt}
=\frac{2}{\kappa^2}\dot H + {\dot\phi}^2 -
8a\frac{d}{dt}\left(\frac{H^2}{a}\frac{d\xi_1(\phi(t))}{dt}\right)\ .
\ee
Eq.~(\ref{GBany7}) can be solved with respect to $\xi_1(\phi(t))$, as
\be
\label{GBany8}
\xi_1(\phi(t))=\frac{1}{8}\int^t dt_1 \frac{a(t_1)}{H(t_1)^2} W(t_1)\ ,\quad
W(t)\equiv \int^{t}
\frac{dt_1}{a(t_1)}
\left(\frac{2}{\kappa^2}\dot H (t_1) + {\dot\phi(t_1)}^2 \right)\ .
\ee
Combining (\ref{GBany4}) and (\ref{GBany8}), the scalar potential $V(\phi(t))$ is:
\be
\label{GBany9}
V(\phi(t)) = \frac{3}{\kappa^2}H(t)^2 - \frac{1}{2}{\dot\phi (t)}^2 - 3a(t)
H(t) W(t)\ .
\ee
We now identify $t$ with $f(\phi)$ and $H$ with $g'(t)$,
where $f$ and $g$
are some functions. Such identification has a close analogy with
the method suggested for the reconstruction of the
scalar-tensor theory in Ref.~\cite{e}.
Let us consider the model where $V(\phi)$ and $\xi_1(\phi)$ can be
expressed in terms of two functions $f$ and $g$ as
\bea
\label{GBany10b}
V(\phi) &=& \frac{3}{\kappa^2}g'\left(f(\phi)\right)^2 - \frac{1}{2f'(\phi)^2}
- 3g'\left(f(\phi)\right) \e^{g\left(f(\phi)\right)} U(\phi) \, \nn
\xi_1(\phi) &=& \frac{1}{8}\int^\phi d\phi_1 \frac{f'(\phi_1)
\e^{g\left(f(\phi_1)\right)} }{g'\left(f(\phi_1)\right)^2} U(\phi_1)\ ,\nn
U(\phi) &\equiv& \int^\phi d\phi_1 f'(\phi_1 ) \e^{-g\left(f(\phi_1)\right)}
\left(\frac{2}{\kappa^2}g''\left(f(\phi_1)\right) + \frac{1}{f'(\phi_1 )^2} \right)\ .
\eea
By choosing $V(\phi)$ and $\xi_1(\phi)$ as (\ref{GBany10b}), we find
the following solution for Eqs.(\ref{GBany4}) and (\ref{GBany5})
(compare with \cite{e}):
\be
\label{GBany11b}
\phi=f^{-1}(t)\quad \left(t=f(\phi)\right)\ ,\quad
a=a_0\e^{g(t)}\ \left(H= g'(t)\right)\ .
\ee
One can straightforwardly check that the solution (\ref{GBany11b})
satisfies the field equation (\ref{GBany6}) \cite{sami}.

%%%%% 
One should now suggest some realistic ansatze for scale factor and scalar.
Our choice below is motivated by the possibility to realize the matter dominated stage as well as cosmic acceleration for such scale factor. The choice of scalar is dictated by the consistency with scale factor 
and simplicity condition. Nevertheless, still it remains some arbytraryness in our choice so several examples will be 
discussed.
%%%%%%%%%%

Consider now as an example the metric
\be
\label{LGB1}
\e^{g(t)}=\left(\frac{t}{t_0}\right)^{g_1}\e^{g_0 t}\ , \quad \phi=f^{-1}(t)=\phi_0 \ln \frac{t}{t_0}\ ,
\ee
where $t_0$, $g_0$, $g_1$, and $\phi_0$ are constants. One further
chooses
\be
\label{LGB2}
\phi_0^2=\frac{2g_1}{\kappa^2}\ .
\ee
Then $U$ is a constant $U=U_0$, and
\bea
\label{LGB3}
V(\phi) &=& \frac{3}{\kappa^2}\left( g_0 + \frac{g_1}{t_0}\e^{- \phi/\phi_0}\right)^2 - \frac{g_1}{\kappa^2t_0^2} \e^{- 2\phi/\phi_0}
- 3U_0\left(g_0 + \frac{g_1}{t_0}\e^{- \phi/\phi_0}\right)\e^{g_1 \phi/\phi_0}\e^{g_0 t_0\e^{\phi/\phi_0}}\ ,\nn
\xi_1 (\phi) &=& \frac{U_0}{8}\int^{t_0 \e^{\phi/\phi_0}} dt_1 \left( g_0 + \frac{g_1}{t_1}\right)^{-2}
\left(\frac{t}{t_0}\right)\e^{g_0 t}\ .
\eea
Eq.~(\ref{LGB1}) leads to
\be
\label{LGB4}
H=g_0 + \frac{g_1}{t}\ .
\ee
Hence, when $t$ is small, the second term in (\ref{LGB4}) dominates
and the scale factor behaves as $a\sim t^{g_1}$. Therefore, if
$g_1=2/3$, a matter-dominated period, where a scalar may be
identified with matter, could be realized.
On the other hand, when $t$ is large, the first term in (\ref{LGB4})
dominates and the Hubble rate $H$ becomes constant.
Therefore, the universe is asymptotically de Sitter space, which is an
accelerating universe.
The three-year WMAP data are analyzed in Ref.~\cite{Spergel}, which
shows that the combined analysis of WMAP with the supernova Legacy
survey (SNLS) constrains the dark energy equation of state $w_{DE}$
pushing it clearly towards the cosmological constant value.
The marginalized best
fit values of the equation of state parameter at 68$\%$ confidence
level are given by $-1.14\leq w_{DE} \leq -0.93$. In case one takes
as a prior that the universe is flat, the combined data gives
$-1.06 \leq w_{DE} \leq -0.90 $.
As in our model the universe goes asymptotically to de Sitter space,
we find $w_{DE} \to -1$.
Therefore, it can easily accommodate these values of $w_{DE}$.
For example, if $g_0 \simeq 40$, $w_{DE}=-0.98$.

In the limit $U_0\to 0$, the Gauss-Bonnet term in (\ref{GBany1})
vanishes and the action (\ref{GBany1})
reduces into that of the usual scalar tensor theory with potential
\be
\label{LGB5}
V(\phi) = \frac{3}{\kappa^2}\left( g_0 + \frac{g_1}{t_0}\e^{- \phi/\phi_0}\right)^2
- \frac{g_1}{\kappa^2t_0^2} \e^{- 2\phi/\phi_0}\ ,
\ee
which reproduces the result in \cite{NOS}.

Let us now consider a second example.
In Einstein gravity with a cosmological constant and matter
characterized by the EOS parameter $w$,
the FRW equation has the following form:
\be
\label{LCDM1}
\frac{3}{\kappa^2}H^2 = \rho_0 a^{-3(1+w)} + \frac{3}{\kappa^2 l^2}\ .
\ee
Here $l$ is the length parameter coming from the cosmological constant.
The solution of (\ref{LCDM1}) is given by
\bea
\label{LCDM2}
a&=&a_0\e^{g(t)}\ ,\nn
g(t)&=&\frac{2}{3(1+w)}\ln \left(\alpha \sinh \left(\frac{3(1+w)}{2l}\left(t - t_0 \right)\right)\right)\ .
\eea
Here $t_0$ is a constant of the integration and
\be
\label{LCDM3}
\alpha^2\equiv \frac{1}{3}\kappa^2 l^2 \rho_0 a_0^{-3(1+w)}\ .
\ee
We now reconstruct the scalar-Gauss-Bonnet gravity model
reproducing (\ref{LCDM2}). If a function $g(t)$ is given by
(\ref{LCDM2}) and $f(\phi)$ is given by
\be
\label{LGB6}
f(\phi) = t_0 - \frac{2l}{3(1+w)}\ln\tanh \left( - \frac{\kappa\sqrt{3(1+w)}}{4}\phi \right)\ ,
\ee
$U(\phi)$ in (\ref{GBany10b}) becomes a constant again, $U=U_0$.
Then, $V(\phi)$ and $\xi(\phi)$ are found to be
\bea
\label{LGB7}
V(\phi)&=& \frac{3}{\kappa^2 l^2} \cosh^2\left(\frac{\kappa\sqrt{3(1+w)}}{2}\phi \right)
- \frac{3(1+w)}{2l^2 \kappa^2} \sinh^2\left(\frac{\kappa\sqrt{3(1+w)}}{2}\phi \right) \nn
&& - \frac{3U_0}{l}\cosh\left(\frac{\kappa\sqrt{3(1+w)}}{2}\phi \right)\left\{ - \frac{1}{\alpha}
\sinh\left(\frac{\kappa\sqrt{3(1+w)}}{2}\phi \right)\right\}^{-2/\left(3(1+w)\right)}\ ,\nn
\xi_1(\phi)&=& - \frac{\alpha U_0 l^3 \kappa}{8\sqrt{3(1+w)}}\int^\phi d\phi_1
\cosh^{-2}\left(\frac{\kappa\sqrt{3(1+w)}}{2}\phi_1 \right)\left\{ - \frac{1}{\alpha}
\sinh\left(\frac{\kappa\sqrt{3(1+w)}}{2}\phi_1 \right)\right\}^{-2/\left(3(1+w)\right) - 1}\ ,
\eea
which again reproduces the result in \cite{NOS} in the limit of
$U_0\to 0$.
Thus, the scalar-Gauss-Bonnet gravity with the scalar potentials under
considerations reproduces the exact $\Lambda$CDM cosmology.

Eq.~(\ref{LCDM2}) shows that when $t\sim t_0$, the scale factor
behaves as $a\sim \left(t - t_0\right)^{2/(3(1+w))}$.
Therefore, if $w=0$, the matter-dominated period can be reproduced.
On the other hand, when $t\to \infty$, $a$ behaves as $a\sim \e^{t/l}$,
which tells us that the universe goes asymptotically to de Sitter
space, with $w_{\rm DE}\to -1$. Therefore, it could be consistent
with WMAP and also with the combined data.

%%%%%%%%%%%%%%%%%%
The Gauss-Bonnet term is usually induced in low-energy string theory. 
Hence, the string theories could determine the form of the functions $V(\phi)$ 
and $\xi_1(\phi)$. In fact, various types of string compactification give the exponential 
potentials $V(\phi)$ and the exponential function for $\xi_1(\phi)$. Then the functions $V(\phi)$ 
and $\xi_1(\phi)$ given in this section, if they are rather simple, could be given by 
such compactified string theories. 
At present, we do not know the theory which could give rather complicated functions $V(\phi)$ 
and $\xi_1(\phi)$ as in (\ref{LGB7}). We cannot, however, exclude such models, which might appear 
in future. Such functions might be related with non-perturbative string effects. 
In our formulation, the universe expansion history dictates the possible forms of $V(\phi)$ 
and $\xi_1(\phi)$, which may be obtained from (some) string compactifications and the non-perturbative 
effects.
%%%%%%%%%%%%%%%%%

\section{$F(G)$ gravity reconstruction from the universe expansion
history and its stability}

We can extend the formulation in the previous section
to $F(G)$ gravity \cite{fGB}, whose action is given by
\be
\label{fG1}
S=\int d^4 x \sqrt{-g}\left[ \frac{R}{2\kappa^2} + F(G) \right]
\ee
The above action can be rewritten by introducing
the auxilliary scalar field $\phi$ as \cite{sami}
\be
\label{fG2}
S=\int d^4 x \sqrt{-g}\left[ \frac{R}{2\kappa^2} - V(\phi) - \xi_1(\phi) G \right]\ .
\ee
By variation over $\phi$, one obtains
\be
\label{fG3}
0=V'(\phi) + \xi_1'(\phi) G\ ,
\ee
which could be solved with respect to $\phi$ as
\be
\label{fG4}
\phi= \phi(G)\ .
\ee
By substituting the expression (\ref{fG4}) into the action
(\ref{fG2}), we obtain the action of $F(G)$ gravity, with
\be
\label{fG5}
F(G)= - V\left(\phi(G)\right) + \xi_1\left(\phi(G)\right)G\ .
\ee
Note that the action (\ref{fG2}) can also be obtained by dropping
the kinetic term of $\phi$ from the action (\ref{GBany1}).

Assuming a spatially-flat FRW universe and the scalar field $\phi$
to depend only on $t$,
we obtain the field equations corresponding to (\ref{GBany4}) and
(\ref{GBany5}):
\bea
\label{fG6}
0&=& - \frac{3}{\kappa^2}H^2 + V(\phi) + 24 H^3 \frac{d \xi_1(\phi(t))}{dt}\ ,\\
\label{fG7}
0&=& \frac{1}{\kappa^2}\left(2\dot H + 3 H^2 \right) - V(\phi)
- 8H^2 \frac{d^2 \xi_1(\phi(t))}{dt^2} - 16H \dot H
\frac{d\xi_1(\phi(t))}{dt} - 16 H^3 \frac{d \xi_1(\phi(t))}{dt}\ .
\eea
Combining the above equations, one gets
\be
\label{fG8}
0=\frac{2}{\kappa^2}\dot H - 8H^2 \frac{d^2
\xi_1(\phi(t))}{dt^2} - 16 H\dot H \frac{d\xi_1(\phi(t))}{dt} + 8H^3 \frac{d\xi_1(\phi(t))}{dt}
=\frac{2}{\kappa^2}\dot H - 8a\frac{d}{dt}\left(\frac{H^2}{a}\frac{d\xi_1(\phi(t))}{dt}\right)\ ,
\ee
which can be solved with respect to $\xi_1(\phi(t))$ as
\be
\label{fG9}
\xi_1(\phi(t))=\frac{1}{8}\int^t dt_1 \frac{a(t_1)}{H(t_1)^2} W(t_1)\ ,\quad
W(t)\equiv \frac{2}{\kappa^2} \int^{t} \frac{dt_1}{a(t_1)} \dot H (t_1)\ .
\ee
Combining (\ref{fG6}) and (\ref{fG9}), the expression
for $V(\phi(t))$ follows:
\be
\label{fG10}
V(\phi(t)) = \frac{3}{\kappa^2}H(t)^2 - 3a(t) H(t) W(t)\ .
\ee
As there is a freedom of redefinition of the scalar field $\phi$
(which corresponds to the choice $f(\phi)=\phi$ in the
last section), we may identify $t$ with $\phi$.
Hence, we consider the model where $V(\phi)$ and $\xi_1(\phi)$ can be
expressed in terms of a single function $g$ as
\bea
\label{fG11}
V(\phi) &=& \frac{3}{\kappa^2}g'\left(\phi\right)^2 - 3g'\left(\phi\right) \e^{g\left(\phi\right)} U(\phi) \ , \nn
\xi_1(\phi) &=& \frac{1}{8}\int^\phi d\phi_1 \frac{\e^{g\left(\phi_1\right)} }{g'(\phi_1)^2} U(\phi_1)\ ,\nn
U(\phi) &\equiv& \frac{2}{\kappa^2}\int^\phi d\phi_1 \e^{-g\left(\phi_1\right)} g''\left(\phi_1\right) \ .
\eea
By choosing $V(\phi)$ and $\xi_1(\phi)$ as (\ref{fG11}), one can easily
find
the following solution for Eqs.(\ref{fG6}) and (\ref{fG7}):
\be
\label{fGB12}
a=a_0\e^{g(t)}\ \left(H= g'(t)\right)\ .
\ee
After that, one can reconstruct $F(G)$ gravity in a way very similar to
the scalar-Gauss-Bonnet theory in the last section.

Although the above formulation is very similar to that in
the scalar-Gauss-Bonnet theory, there could be some difference:
in scalar-GB gravity, since the scalar field has a kinetic
term, it can propagate and there can be the possibility to generate an
extra force besides the Newtonian one. On the other hand,
$F(G)$ gravity has no kinetic
term for the scalar field and an extra force cannot be generated.
In fact, one can consider the perturbation around the de Sitter
background (compare with section five), by writing the metric as
$g_{\mu\nu}=g_{(0)\mu\nu} + h_{\mu\nu}$.
Here, the Riemann tensor in the de Sitter background is given by
\be
\label{GB35}
R_{(0)\mu\nu\rho\sigma}=H_0^2\left(g_{(0)\mu\rho}g_{(0)\nu\sigma}
- g_{(0)\mu\sigma}g_{(0)\nu\rho}\right)\ .
\ee
The flat background corresponds to the limit $H_0\to 0$.
For simplicity, if we choose the gauge conditions
$g_{(0)}^{\mu\nu} h_{\mu\nu}=\nabla_{(0)}^\mu h_{\mu\nu}=0$,
we find from the equation of motion without the energy-momentum tensor,
\be
\label{GB38}
0=\frac{1}{4\kappa^2} \left( \nabla^2 h_{\mu\nu} - 2H_0^2 h_{\mu\nu}\right)\ .
\ee
Since the contribution form the Gauss-Bonnet term does not appear
except in the length parameter $1/H_0$ of the de Sitter space, the
only propagating mode should
be the graviton in the $F(G)$ gravity.

Let us now investigate the stability of cosmological solutions in
$F(G)$ gravity. The stability study of such theory on a de Sitter
background will be presented in seventh section where constraints to
$F(G)$ from stability condition will be defined.
For $V(\phi)$ and $\xi_1(\phi)$ given by (\ref{fG11}), Eqs.(\ref{fG3})
and (\ref{fG6}) yield
\be
\label{Ins1}
\dot H = - H^2 + \frac{g'(\phi)^2}{H^2}\left(g''(\phi) + g'(\phi)^2\right)\ ,\quad
\dot\phi = \frac{g'(\phi)^3}{H^3} + \frac{g'(\phi)^2}{\kappa^2 H^3}\e^{-g(\phi)}
\left(H^2 - g'(\phi)\right)\ .
\ee
We now consider the perturbation of solution (\ref{fGB12}), that is,
\be
\label{Ins2}
H=g'(t)\ ,\quad \phi=t\ ,
\ee
as
\be
\label{Ins3}
H=g'(t) + \delta H\ ,\quad \phi=t + \delta \phi\ .
\ee
Here it is asuumed that $\delta H$ and $\delta \phi$ only depend
on time $t$. From (\ref{Ins1}), we obtain
\bea
\label{Ins3b}
&& \frac{d}{dt}\left(\begin{array}{c} \delta H \\ \delta \phi \end{array} \right)
=M\left(\begin{array}{c} \delta H \\ \delta \phi \end{array} \right)\ ,\nn
&& M = \left(\begin{array}{cc} - \frac{2g''(t)}{g'(t)} - 4g'(t) & g'''(t) + 4g'(t) g'' (t) + \frac{2g''(t)^2}{g'(t)} \\
-\frac{3}{g'(t)} + \frac{2\e^{-g(t)}}{\kappa^2} & - \left(-\frac{3}{g'(t)} + \frac{2\e^{-g(t)}}{\kappa^2}\right)g''(t)
\end{array}\right)\ .
\eea
If the two real parts of both eigenvalues are either negative or
vanish, the system turns out to be stable under the perturbation.
If $\lambda$ is the eigenvalue, the corresponding eigenvalue equations
are written as
\be
\label{Ins4}
0 =\lambda^2 - \left( - \frac{g''(t)}{g'(t)} - 4g'(t) - \frac{2g''(t)\e^{-g(t)}}{\kappa^2}\right)\lambda
+ \left( \frac{3}{g'(t)} - \frac{2\e^{-g(t)}}{\kappa^2} \right)g'''(t)\ ,
\ee
whose determinant is given by
\be
\label{Ins5}
D=\left( - \frac{g''(t)}{g'(t)} - 4g'(t) - \frac{2g''(t)\e^{-g(t)}}{\kappa^2}\right)^2
- 4 \left( \frac{3}{g'(t)} - \frac{2\e^{-g(t)}}{\kappa^2} \right)g'''(t)\ .
\ee
The solutions of (\ref{Ins4}) are
\be
\label{Ins6}
\lambda = \lambda_\pm \equiv \frac{1}{2}\left( - \frac{g''(t)}{g'(t)} - 4g'(t) - \frac{2g''(t)\e^{-g(t)}}{\kappa^2}
\pm \sqrt{D} \right)\ .
\ee
When $D<0$, the eigenvalues are complex and, in order that the real
parts are negative (as required for stability), we have
\be
\label{ins7}
A(t)\equiv - \frac{g''(t)}{g'(t)} - 4g'(t) - \frac{2g''(t)\e^{-g(t)}}{\kappa^2} <0\ .
\ee
When $D>0$, both eigenvalues are negative and the system is stable
if and only if
\be
\label{ins8}
A(t) = - \frac{g''(t)}{g'(t)} - 4g'(t) - \frac{2g''(t)\e^{-g(t)}}{\kappa^2} < 0\ ,\quad
B(t) \equiv \left( \frac{3}{g'(t)} - \frac{2\e^{-g(t)}}{\kappa^2} \right)g'''(t)>0\ .
\ee

One may investigate the stability of the cosmological solution (\ref{LGB4}):
\be
\label{ins9}
g(t)=\int dt\,H=g_0t + g_1\ln\frac{t}{t_0}\ .
\ee
Here $t_0$ is a constant of the integration. Let us consider the case
that $g_1=2/3$, which corresponds to dust.
In the early universe, where $t\to 0$, we find
\be
\label{ins10}
A(t)\sim \frac{4t_0^{2/3}}{3\kappa^2}t^{-8/3}>0\ .
\ee
Hence, the solution is unstable, which is the typical property of the
early-time universe.
On the other hand, in the late universe, where $t\to\infty$, one finds
\be
\label{ins11}
A(t)\sim - 4g_0<0\ ,\quad B(t)\sim \frac{6}{g_0t^2}>0\ .
\ee
Thus, the solution is stable and the future universe does not enter
into any Big Rip-like era\cite{brett}.

We also investigate the stability of the $\Lambda$CDM cosmology
corresponding to (\ref{LCDM2}).
It is convenient to consider the case that $w=0$, which corresponds to
dust.
Hence, in the early universe, where $t\to t_0$, we find
\be
\label{ins12}
A(t)\sim \frac{4}{3\kappa^2}\left(\frac{2l}{3\alpha}\right)^{2/3}\left(t-t_0\right)^{-8/3}>0\ .
\ee
Therefore, the solution is unstable.
On the other hand, in the late universe, it follows that
\be
\label{ins13}
A\sim - \frac{4}{l}<0\ ,\quad B(t)\sim \frac{54}{l^2}\e^{-3\left(t - t_0\right)/l}>0\ .
\ee
Thus, the solution is stable again.

In general, one can consider the case when $g(t)$ behaves
asymptotically (that is, in either the early or the late universe)
as
\be
\label{ins14}
g(t)=h_0\ln \frac{t}{t_0}\ .
\ee
The case $h_0=2/3$ corresponds to dust dominated universe. For
simplicity, we only consider the case that
$h_0>0$ (non-phantom case). Then, we find
\be
\label{ins15}
A(t)=\frac{1-4h_0}{t} + \frac{2h_0 t_0^{h_0}}{\kappa^2}t^{-h_0 - 2}\ ,\quad
B(t)=\left(\frac{3}{h_0}t - \frac{2}{\kappa^2}\left(\frac{t}{t_0}\right)^{-h_0}\right)\frac{2h_0}{t^3} \ .
\ee
When $0<h_0<1/4$, it always follows that $A(t)>0$; therefore, the solution is unstable.
In the early universe, where $t\to 0$, $A(t)$ behaves as
\be
\label{ins16}
A(t)\sim \frac{2h_0}{\kappa^2}\left(\frac{t}{t_0}\right)^{-h_0}\frac{1}{t^2}>0\ .
\ee
As a consequence the solution is always unstable in the early universe.
%%%%%%%%%%%%%%%%
The early universe instability indicates that
the simple solution (\ref{ins9}) is unstable and the inflationary epoch
should end as is widely expected.
%%%%%%%%%%%%%%%%

On the other hand, in the late universe, where $t\to\infty$, we find
\be
\label{ins17}
A(t)\sim \frac{1-4h_0}{t}\ ,\quad B(t)\sim \frac{6}{t^2}>0\ .
\ee
Then as long as $h_0>1/4$, the solution should be stable.

We have thus presented several examples of the reconstruction
program of modified
Gauss-Bonnet gravity from the universe expansion history which include
the epochs
of early-time inflation, the matter dominated era,
the deceleration-acceleration transition, and the acceleration
epoch. The stability conditions of such $\Lambda$CDM-like cosmology
have been investigated, showing that the late universe can indeed be
stable, without exhibiting any type of Big Rip behavior (for their
classification, see \cite{tsujikawa}). In a similar way, one can
consider dark energy cosmology in other versions of the modified
Gauss-Bonnet theory \cite{fGB,GB,emili} and relate them with the
radiation and matter-dominated epochs.

\section{Reconstruction of modified Gauss-Bonnet gravity with matter}

It is not difficult to extend the above formulation in scalar-Gauss-Bonnet gravity and $F(G)$ gravity to include several matter terms
with constant equation of state (EoS) parameters $w_i\equiv p_i
/ \rho_i$.
Here $\rho_i$ and $p_i$ are the energy density and pressure of the
$i$-th matter term. We now include several kinds of matter, like
radiation, baryons, cold dark matter, etc.
Then, instead of (\ref{GBany4}) and (\ref{GBany5}) or (\ref{fG6})
and (\ref{fG7}), the corresponding FRW equations are given as
\bea
\label{GBany16}
0&=& - \frac{3}{\kappa^2}H^2 + \frac{\eta}{2}{\dot\phi}^2 + V(\phi) + \sum_i \rho_i
+ 24 H^3 \frac{d \xi_1(\phi(t))}{dt}\ ,\\
\label{GBany17}
0&=& \frac{1}{\kappa^2}\left(2\dot H + 3 H^2 \right) + \frac{\eta}{2}{\dot\phi}^2
- V(\phi) + \sum_i p_i - 8H^2 \frac{d^2 \xi_1(\phi(t))}{dt^2} - 16H \dot H \frac{d\xi_1(\phi(t))}{dt}
- 16 H^3 \frac{d \xi_1(\phi(t))}{dt}\ .
\eea
Here $\eta=1$ corresponds to scalar-Gauss-Bonnet gravity and
$\eta=0$ to $F(G)$ gravity in the form (\ref{fG2}).
The energy conservation law
\be
\label{GBany18}
\dot\rho_i + 3H\left(\rho_i + p_i\right)=0\ ,
\ee
gives
\be
\label{GBany19}
\rho_i=\rho_{i0} a^{-3(1+w_i)}\ ,
\ee
with a constant $\rho_{i0}$.
Instead of (\ref{GBany10b}), one should consider the model with
\bea
\label{GBany20b}
V(\phi) &=& \frac{3}{\kappa^2}g'\left(f(\phi)\right)^2 - \frac{\eta}{2f'(\phi)^2}
- 3g'\left(f(\phi)\right) \e^{g\left(f(\phi)\right)} U_m(\phi) \ ,\nn
\xi_1(\phi) &=& \frac{1}{8}\int^\phi d\phi_1 \frac{f'(\phi_1)
\e^{g\left(f(\phi_1)\right)}
U_m(\phi_1)}{g'\left(f(\phi_1)\right)^2}\ ,\nn U_m(\phi)&\equiv &
\int^\phi d\phi_1 f'(\phi_1)\e^{-g\left(f(\phi_1)\right)}
\left(\frac{2}{\kappa^2}g''\left(f(\phi_1)\right) +
\frac{\eta}{2f'(\phi_1 )^2} + \sum_ i (1+w_i)\rho_{i0}
a_0\e^{-3(1+w_i)g\left(f(\phi_1)\right)}\right) \ .
\eea
In this way we re-obtain the solution (\ref{GBany11b}), also in the case when
matter is included. This expression is different from
(\ref{GBany10b}) due to the last terms in $U_m$, that is, the terms
proportional to $\rho_{i0}$.
%%%%%%%%%%%%%
This term appears since we include matter. 
In (\ref{GBany10b}), even without matter, the transition
from the matter dominated phase to acceleration can occur. 
Therefore in order that the transition could occur, we need not alway the matters themselves. 
In the real universe, of course, there are matters as in the model in this section. 
If we consider the model where $V(\phi)$ and $\xi_1(\phi)$ are given in (\ref{GBany20b}) but matters 
are {\it not} included, the matter dominated phase would not appear. Then in the model (\ref{GBany20b}) 
{\it with} matter, the matter dominated phase to the acceleration phase occurs not only due to pure
Gauss-Bonnet term effects but due to combined effect of Gauss-Bonnet term
and matter presence.
%%%%%%%%%%%

In case of $F(G)$ gravity with $\eta=0$, we can choose $f(\phi)=\phi$.
Then, expressions (\ref{GBany20b}) can be simplified:
\bea
\label{fGB13}
V(\phi) &=& \frac{3}{\kappa^2}g'\left(\phi\right)^2 - 3g'\left(\phi\right) \e^{g\left(\phi\right)}
\tilde U_m(\phi) \ ,\nn
\xi_1(\phi) &=& \frac{1}{8}\int^\phi d\phi_1 \frac{\e^{g\left(\phi_1\right)} \tilde U_m(\phi_1)}{g'(\phi_1)^2}\ ,\nn
\tilde U_m(\phi)&\equiv & \int^\phi d\phi_1 \e^{-g\left(\phi_1\right)}
\left(\frac{2}{\kappa^2}g''\left(\phi_1\right) + \sum_i (1+w_i)\rho_{0i} a_0\e^{-3(1+w_i)g\left(\phi_1\right)}\right) \ .
\eea

It is instructive to consider an example of $F(G)$ gravity with only dust as matter,
which could be baryons and dark matter with $w=0$, and $g$ given by
\be
\label{ex1}
g(\phi)=\frac{2}{3}\ln\left(\sinh\left(C\phi\right)\right)\ ,\quad
C\equiv \frac{2}{3}\sqrt{\frac{\rho_{0d} a_0 \kappa^2}{3}}\ .
\ee
This $g(\phi)$ corresponds to (\ref{LCDM2}) with only dust: $w=0$.
Here $g(\phi)$ is written in a little bit different way as will soon
be seen.
Eq.~(\ref{ex1}) shows that $U$ is a constant $U=U_0$, and
\bea
\label{ex2}
V(\phi) &=& \frac{2C^2}{\kappa^2}\coth^2 (C\phi) - 3CU_0 \coth (C\phi) \sinh^{2/3} (C\phi)\ ,\nn
\xi_1 &=& \frac{U_0}{8}\int^\phi d\phi_1 \sinh^{-4/3} (C\phi) \cosh^2 (C\phi)\ .
\eea
Eq.~(\ref{ex1}) indicates that, when $\phi=t$ is small, $g(\phi)$ behaves as
$g(\phi)\sim (2/3)\ln \phi$
and, therefore, the Hubble rate behaves as
$H(t)=g'(t)\sim (2/3)/t$, which surely reproduces the matter dominated
phase. On the other hand, when $\phi=t$ is large,
$g(\phi)$ behaves as $g\sim (2/3) (C\phi)$, that is, $H \sim 2C/3$ and the universe asymptotically goes to de Sitter
space. Therefore, the model given by (\ref{ex2}) {\it with matter} shows
the transition from the matter dominated phase
to the acceleration universe, which is asymptotically de Sitter space.
In fact, by comparing (\ref{ex1}) with (\ref{LCDM2}), one can identify
\be
\label{ex3}
\alpha=1\ ,\quad C=\frac{3(1+w)}{2l}\ .
\ee
We now check if the transition from the matter dominated phase to
the acceleration phase could occur or not in this model
{\it without matter}.
It can be shown that, if one assumes the existence of the
matter dominated phase, a contradiction appears.
Note that in this model without matter, there is no reason to identify
$\phi$ with $t$.
Eq.~(\ref{fG3}) gives
\be
\label{ex4}
0= - \frac{4C^2}{\kappa^2}\frac{\cosh (C\phi)}{\sinh^3 (C\phi)} + B^2U_0 \sinh^{-4/3} (C\phi)
\left(3 - 2\cosh^2 (C\phi) \right) + \frac{U_0}{8}\sinh^{-4/3} (C\phi) \cosh^2 (C\phi)\, G \ .
\ee
In the matter dominated phase, the curvature $R$ and, therefore, the
Gauss-Bonnet invariant should be large.
Then Eq.~(\ref{ex4}) tells us that $\phi$ should be small in the matter dominated phase, and we find
\be
\label{ex5}
G\sim \frac{32C^{1/3}}{\kappa^2 U_0}\phi^{-5/3}\ .
\ee
On the other hand, in the matter dominated phase,
the Hubble rate $H(t)$ behaves as $H\sim (2/3)/t$ and, therefore,
\be
\label{C5}
G=24 \left(H^2 \dot H + H^4\right) \sim - \frac{64}{27t^4}\ .
\ee
Comparing (\ref{C5}) with (\ref{ex5}), it follows that
$\phi\sim t^{12/5}$. Hence,
\be
\label{ex6}
V(\phi)\sim \frac{2C^2}{\kappa^2\phi^2}\sim t^{-24/5}\ ,\quad
\xi_1\sim - \frac{3U_0}{8B^{4/3}}\phi^{-1/3}\sim t^{-4/5}\ .
\ee
This behavior is in conflict with (\ref{fG6}). The first equation
(\ref{fG6}) may be written as the usual FRW equation with an
inhomogeneous EoS ideal fluid \cite{inh} (for a particular example
of such an
inhomogeneous EoS ideal fluid interpreted as time-dependent bulk
viscosity, see \cite{brevik})
\be
\label{ex7}
\frac{3}{\kappa^2}H^2 =\rho_G\ ,\quad \rho_G\equiv V(\phi) + 24 H^3 \frac{d \xi_1(\phi(t))}{dt}\ .
\ee
Eqs.(\ref{C5}) and (\ref{ex6}) show that $\rho_G$ behaves as
$t^{-24/5}$, but $H^2$ behaves as $t^{-2}$.
Therefore, there is a discrepancy between the power of $t$ on both
sides of (\ref{ex7}).
Thus, the matter dominated phase can in no way be realized {\it without}
matter, and consequently the model (\ref{ex2})
does {\it not} generate the transition from the matter dominated phase to the acceleration phase {\it without} matter.

As the EoS parameter of the dark energy
in the present universe is almost $-1$, the universe could
approach the de Sitter space asymptotically. Let us consider the form
of the action when the universe does become de Sitter space.
We now assume
\be
\label{LGB8}
g(t)=H_0 t\ ,\quad f(\phi)=f_0 \phi\ .
\ee
Here $H_0$ and $f_0$ are constant. Since the Hubble parameter $H$
is given by $H=\dot g=H_0$, the universe is in fact
de Sitter space. For $F(G)$ gravity, one may put $f_0=1$.
It follows that
\bea
\label{LGB9}
U_m(\phi)&=& - \frac{\eta}{2H_0 f_0^2}\e^{-H_0f_0\phi}
+ \sum_i \frac{(1+w_i)\rho_{i0} a_0}{(2+3w_0)H_0}\e^{\left(2+3w_i\right)H_0 f_0 \phi} + U_0\ ,\nn
V(\phi) &=& \frac{3H_0^2}{\kappa^2} + \frac{\eta}{f_0^2} - \sum_i \frac{3(1+w_i)\rho_{i0} a_0}{2+3 w_i}\e^{3(1+w_i)H_0 f_0 \phi}
- 3H_0 U_0 \e^{H_0 f_0 \phi}\ , \nn
\xi_1(\phi) &=& - \frac{\eta\phi}{16H_0^3 f_0}
+ \sum_i \frac{(1+w_i)\rho_{i0} a_0}{24\left(1+w_i\right)\left(2+3 w_i\right)H_0^4}\e^{3(1+w_i)H_0 f_0 \phi}
+ \frac{U_0}{8H_0^3}\e^{H_0 f_0 \phi} + \xi_0\ .
\eea
Here $U_0$ and $\xi_0$ are integration constants. Effectively
$\xi_0=0$, since a constant times the GB term is
a total derivative. Even in the presence of matter, the de Sitter
universe could still be realized by adding an exponential potential and
the GB coupling.

It is interesting to consider the effect of matter to the
corresponding cosmology
(\ref{LCDM2}). This
could be done by adding the part compensating
the contribution from matter to $U(\phi)$, which is chosen to be a constant in the model (\ref{LCDM2}), as
\bea
\label{LGB10}
U_m(\phi) &=& U_0 + \sum_i U_i(\phi)\ , \nn
U_i(\phi) &\equiv& \left(1+w_i\right)\rho_{i0}a_0
\int^{f(\phi)} dt \left\{\alpha\sinh\left(\frac{3(1+w)}{2l}\left(t - t_0\right) \right)\right\}^{- (4 + 3 w_i)/3(1+w)}\ .
\eea
Then $V(\phi)$ and $\xi_1(\phi)$ (\ref{LGB7}) are modified as
\bea
\label{LGB11}
V(\phi)&=& \frac{3}{\kappa^2 l^2} \cosh^2\left(\frac{\kappa\sqrt{3(1+w)}}{2}\phi \right)
- \frac{3(1+w)}{2l^2 \kappa^2} \sinh^2\left(\frac{\kappa\sqrt{3(1+w)}}{2}\phi \right) \nn
&& - \frac{3}{l}\left( U_0 + \sum_i U_i(\phi)\right)
\cosh\left(\frac{\kappa\sqrt{3(1+w)}}{2}\phi \right)\left\{ - \frac{1}{\alpha}
\sinh\left(\frac{\kappa\sqrt{3(1+w)}}{2}\phi \right)\right\}^{-2/\left(3(1+w)\right)}\ , \nn
\xi_1(\phi)&=& - \frac{\alpha l^3 \kappa}{8\sqrt{3(1+w)}}\int^\phi d\phi_1
\left( U_0 + \sum_i U_i(\phi_1) \right) \nn
&& \times \cosh^{-2}\left(\frac{\kappa\sqrt{3(1+w)}}{2}\phi_1 \right)\left\{ - \frac{1}{\alpha}
\sinh\left(\frac{\kappa\sqrt{3(1+w)}}{2}\phi_1 \right)\right\}^{-2/\left(3(1+w)\right) - 1}\ ,
\eea
Therefore, there exists a scalar-GB gravity which shows the
transition from the matter dominated phase to the acceleration phase,
even in the presence of matter.
This theory could be regarded as an extension of (\ref{ex1}) with only
dust (baryon and/or dark matter).
Thus, in the model (\ref{LGB10}) without matter, the matter
dominated phase cannot be realized
and therefore the transition from the matter
dominated phase to the acceleration era does not occur.

Let us investigate the asymptotic behavior of $U_i(\phi)$ in
(\ref{LGB10}).
When $t=f(\phi)\to t_0$, one gets
\be
\label{LGB12}
U_i(\phi)\to \frac{2l}{\alpha\left\{ - 1 + 3 (w - w_i)\right\}}\left\{\frac{3\alpha(1+w)}{2l}
\left(f(\phi) - t_0\right)\right\}^{\left\{-1 + 3 (w - w_i)\right\}/3(1+w)} \ .
\ee
When the model containing the matter dominated phase is considered, one
can find $w=0$.
Since usually $w_i\geq 0$, the power $\left[-1 + 3 (w - w_i)\right]/3(1+w)$ could be
always negative. Hence, $U_i(\phi)$ could be singular when
$t=f(\phi)\to t_0$. This is not something
extraordinary. For example, the scale factor of the FRW universe
with matter, whose EoS parameter
is a constant $w$, behaves as $a\sim t^{-\frac{2}{3(1+w)}}$, which is singular at $t=0$.

On the other hand, when $t=f(\phi)\to \infty$, we find
\be
\label{LGB13}
U_i(\phi) \to U_{iF} - \left(\frac{\alpha}{2}\right)^{-(4+3w_i)/3(1+w)} \frac{2l}{4 + 3 w_i}
\exp\left(- \frac{4+3w_i}{2l}\left(f(\phi) - t_0\right)\right)\ .
\ee
Here $U_{iF}$ is an integration constant.
When $t=f(\phi)\to \infty$, the second term vanishes very rapidly.
We may choose $U_{iF}=0$. Thus, in the case $U_0\neq 0$,
the corrections generated by adding matter could be neglected.
These corrections become important
only when $t=f(\phi)\to t_0$, which may correspond to the early universe.

Our study shows the existence of a big class of modified Gauss-Bonnet
gravity models with matter where the cosmological sequence of matter
dominance, deceleration-acceleration transition and cosmic
acceleration occur {\it only} in the presence of matter.

\section{A compensating dark energy}

Some versions of scalar-Gauss-Bonnet or $f(G)$ gravity (with specific
potentials) do not have a matter dominated stage as a solution, not
even in the presence of matter.
Similarly, some of them have a
stable matter dominated era as a solution,
hence, no transition to an acceleration epoch occurs. It is
remarkable that even in this case, a realistic cosmology may emerge
at the price of introducing a compensating dark energy. Such
scenario has been proposed in Ref.~\cite{reconstruction}, based
on the example of modified gravity of Ref.~\cite{prd}. Let us
introduce a
compensating dark energy, which could be an ideal fluid and may help
to realize the matter dominated
and deceleration-acceleration transition phases in modified
GB gravity.
One may question the necessity of such ideal fluid, saying, that already
the Einstein gravity with negative pressure ideal fluid may describe the
acceleration of the universe. The below scenario is more sophisticated: it
is supposed that the acceleration is produced by scalar-GB interaction
and compensating dark energy is negligible in the late epoch.
(Hence, without GB term no cosmic acceleration occurs).
The role of compensating dark energy is relevant only in matter dominated
era: it helps in the realization of matter dominance.

When the energy density $\rho_c$ and $p_c$ of the compensating dark energy
are added, the FRW equations become
\bea
\label{C1}
0&=& - \frac{3}{\kappa^2}H^2 + \frac{\eta}{2}{\dot\phi}^2 + V(\phi) + 24 H^3
\frac{d \xi_1(\phi(t))}{dt} + \rho_c + \rho_m \ ,\\
\label{C2}
0&=& \frac{1}{\kappa^2}\left(2\dot H + 3 H^2 \right) + \frac{\eta}{2}{\dot\phi}^2 - V(\phi)
- 8H^2 \frac{d^2 \xi_1(\phi(t))}{dt^2} - 16H \dot H
\frac{d\xi_1(\phi(t))}{dt} - 16 H^3 \frac{d \xi_1(\phi(t))}{dt}
+ p_c + p_m \ .
\eea
Here, $\rho_m$ and $p_m$ are contributions from matter, which may also
include dark matter.
In the matter dominated phase, the contribution from the scalar-GB
terms could correspondingly be canceled
by that coming from the compensating dark energy.
In other words, in the matter dominated era the following conditions hold
\bea
\label{C3}
\rho_c &\sim & - \left\{\frac{\eta}{2}{\dot\phi}^2 + V(\phi) + 24 H^3 \frac{d \xi_1(\phi(t))}{dt}\right\} \ ,\\
\label{C4}
p_c & \sim & - \left\{\frac{\eta}{2}{\dot\phi}^2 - V(\phi) - 8H^2 \frac{d^2 \xi_1(\phi(t))}{dt^2} - 16H \dot H
\frac{d\xi_1(\phi(t))}{dt} - 16 H^3 \frac{d \xi_1(\phi(t))}{dt} \right\} \ .
\eea
If matter is dust, as baryons or cold dark matter, the Gauss-Bonnet invariant $G$ behaves as (\ref{C5}).

For simplicity, we will first consider $F(G)$ gravity with $\eta=0$.
When $H=h/t$, $G$ is given by $G=24h^3(h-1)/t^4$. Then $G$ changes its sign when $h=1$.
In the decelerating universe, $h<-1$, but in the accelerating
one the sign of $h$ changes: $h>-1$. When the decelerating
universe turns to the accelerating phase, $G$ changes its sign.
In this case, solving (\ref{fG3}), if we do not choose $V(\phi)$ and
$\xi_1$ properly, there is no
solution. In the case $H=h/t$, Eqs.~(\ref{fG11}) yield
\be
\label{C6}
V(\phi)=\frac{3h^2(h-1)}{(h+1) \kappa^2 \phi^2}\ ,\quad
\xi_1(\phi) = \frac{\phi^2}{8h(h+1)\kappa^2}\ .
\ee
We now consider a new model by replacing $h$ in (\ref{C6}) with $h(\phi)$, adiabatically depending on $\phi$,
so that one can neglect all derivatives
$h'(\phi)$, $h''(\phi)$, $\cdots$:
\be
\label{C7}
V(\phi)=\frac{3h(\phi)^2(h(\phi)-1)}{(h(\phi)+1) \kappa^2 \phi^2}\ ,\quad
\xi_1(\phi) = \frac{\phi^2}{8h(\phi)(h(\phi)+1)\kappa^2}\ .
\ee
Using (\ref{fG3}) and neglecting $h'(\phi)$, $h''(\phi)$, $\cdots$, we
find, as expected,
\be
\label{C8}
\phi^2 = \frac{24h(\phi)^3 \left( 1 - h(\phi) \right)}{G}\ .
\ee
As the adiabatic approximation is used, our identification $\phi=t$
can again be introduced.
For the matter dominated era, by using (\ref{C3}) and (\ref{C4}), we find
\be
\label{C9}
\rho_c \sim - \frac{3h(t)^2}{\phi^2} \ ,\quad
p_c \sim \frac{h ( 3h(t) - 2 )}{\kappa^2 \phi^2}\ ,
\ee
which gives the EoS parameter
\be
\label{C10}
w_c = \frac{p_c}{\rho_c} \sim \frac{2-3h}{3h}\ .
\ee
Hence, when $h\to 2/3$, which corresponds to dust, $w_c\to 0$ as
expected.
We should note that the energy density of
the compensating dark energy $\rho_c$ is now negative, which might be
possible if there is a negative cosmological constant (for instance,
the one produced by anti-deSitter space), which effectively shifts the
energy a by negative constant.

We now consider the scalar-Gauss-Bonnet theory (with $\eta=1$) of
Ref.~ \cite{sasaki}, where
\be
\label{C11}
V=V_0\e^{-\frac{2\phi}{\phi_0}}\ ,\quad \xi_1(\phi)=\xi_0 \e^{\frac{2\phi}{\phi_0}} \ .
\ee
Without matter, the theory with those potentials (\ref{C11}) can
be solved \cite{sasaki}
\bea
\label{C12}
& H=\frac{h_0}{t}\ ,\quad \phi=\phi_0 \ln \frac{t}{t_1}\ ,\quad & \left(\mbox{when}\ h_0>0\right) \ , \nn
& H=-\frac{h_0}{t_s - t}\ ,\quad \phi=\phi_0 \ln \frac{t_s - t}{t_1}\ ,\quad & \left(\mbox{when}\ h_0<0\right) \ .
\eea
Here $h_0$ and $t_1$ are related with $V_0$ and $\xi_0$, as
\bea
\label{C13}
&& V_0 t_1^2= - \frac{1}{\kappa^2\left(1 + h_0\right)}\left\{3h_0^2 \left( 1 - h_0\right)
+ \frac{\phi_0^2 \kappa^2 \left( 1 - 5 h_0\right)}{2}\right\}\ ,\nn
&& \frac{48 \xi_0 h_0^2}{t_1^2}= \frac{6}{\kappa^2\left( 1
+ h_0\right)}\left(h_0 - \frac{\phi_0^2 \kappa^2}{2}\right)\ .
\eea
Since the above $h_0$ is a constant, the EoS parameter $w$ is also a constant.
Therefore, in this model without matter and compensating dark energy,
the transition from the matter dominated phase to the acceleration phase could {\it not} be generated.

In the following, it is enough to consider the case with $h_0>0$ only.
Let us now add the contribution from the
compensating dark energy and usual matter, and assume $H(t)=h(t)/t$
with a slowly varying
function $h(t)$. The time dependence of $\phi$ can be determined by solving the scalar
field equation (\ref{GBany6}). Neglecting $h'(\phi)$, $h''(\phi)$,
$\cdots$ in the adiabatic regime,
we get
\be
\label{C14}
\phi=\phi_0\ln \frac{t}{t_1(t)}\ .
\ee
Here $t_1(t)$ is found from the following equation
\be
\label{C15}
0=\left( 1 - 3h(t) \right)\phi_0^2 + 2V_0 t_1(t)^2
- \frac{48 \xi_0 h(t)^3}{t_1^2}\left(h(t) - 1\right)\ .
\ee Hence, in the matter dominated phase, Eqs.~(\ref{C3}) and
(\ref{C4}) are \bea \label{C16}
\rho_c &\sim & - \left\{\frac{\phi_0^2}{2} + V_0 + 48\xi_0 h(t)^3 \right\}\frac{1}{t^2} \ ,\\
\label{C17}
p_c & \sim & - \left\{\frac{\phi_0^2}{2} - V_0 + 16\xi_0h(t)^2 \left( 1 - 2 h(t) \right)\right\}
\frac{1}{t^2} \ ,
\eea
which give the effective EoS parameter
\be
\label{C18}
w_c = \frac{\frac{\phi_0^2}{2} - V_0 + 16\xi_0h(t)^2\left( 1 - 2 h(t)\right)}
{\frac{\phi_0^2}{2} + V_0 + 48\xi_0 h(t)^3}\ .
\ee
Thus, scalar-Gauss-Bonnet gravity with potentials (\ref{C11}) and without
matter and/or compensating dark energy does not contain the
transition from the matter dominated phase to the acceleration phase.
Nevertheless,
adding the matter and the compensating
dark energy (\ref{C16}),
the matter dominated phase could be realized even for the model
(\ref{C11}), similarly to other classes of modified gravity
\cite{reconstruction}.
As expressed in (\ref{C3}), the compensating dark dark energy cancels
the contributions from
the scalar-Gauss-Bonnet term in the early universe and, therefore, the
system effectively reduces to Einstein gravity
coupled with matter, where the matter dominated universe can in fact be
generated. In the late universe, the contributions from
matter and the compensating dark energy become {\it small}
and the accelerated expanding universe naturally takes over.
A similar scenario can be devised to improve the emergence of the
matter dominance stage in other versions of modified gravity and
scalar-tensor theories (for a recent study of accelerated cosmologies
in scalar-tensor theories, see \cite{scalar} and the
references therein).

\section{Discussion}

In summary, we have tried to show in this paper that string-inspired
scalar-Gauss-Bonnet gravity and modified Gauss-Bonnet gravity are
indeed interesting alternatives for dark energy. These theories are
in fact closely related and, what is also important, may have a
stringy origin. Here, the reconstruction program from the universe
expansion history for those theories has been carried out
successfully. Several explicit examples of the same (with some
specific potentials) have been presented where the cosmological
sequence of the matter dominance, deceleration-acceleration
transition and cosmic acceleration occurs very naturally. Moreover,
the accelerated universe can be asymptotically de Sitter or it may
correspond to an exact $\Lambda$CDM cosmology. The study of
perturbations around the cosmological solutions above has been
performed too. There is no problem to include usual matter with
specific equation of state into such consideration. In that case,
one can construct versions of the Gauss-Bonnet gravities above where
a matter dominance period and a deceleration-acceleration transition
occur {\it only} in the presence of matter. It is also remarkable
that, even in the case when such intermediate universe is not a
solution of some modified Gauss-Bonnet gravity, it can actually be
made so, at the price of introducing some compensating dark energy.

%%%%%%%%%%%%%%%%%%
Since the Gauss-Bonnet term is usually induced in low-energy string theory, 
the functions $V(\phi)$ and $\xi_1(\phi)$ could contain the information about the 
string compactification and/or stringy non-perturbative effects. 
By the reconstruction program in this paper, we suggested the possible forms of $V(\phi)$ 
and $\xi_1(\phi)$, which are cosmologically viable. 
%%%%%%%%%%%%%%%%%

Due to the fundamental role of the de Sitter space which appears in
our scenario as the final state of the universe, special attention
has been paid to such space. In particular, the one-loop effective
action of $F(G)$ gravity has been found on the de Sitter background.
This effective action was then used to derive stability criteria for
the modified Gauss-Bonnet gravity theory. Some numerical examples
showing explicit versions of modified Gauss-Bonnet gravity with a
stable de Sitter vacuum have been presented.

The successful reconstruction of string-inspired, Gauss-Bonnet
gravity from the universe expansion history, performed in the
present work, shows that it actually represents a reasonable
gravitational alternative for dark energy. Having in mind the
promising results obtained in the comparison of such a theory with
observational data (see, for example, \cite{Mota}), it becomes clear
that it deserves careful attention. Moreover, the theory
successfully passes the check of the three-year WMAP observational
data. Needless to say, as with any other alternative to General
Relativity at the current and/or future universe, additional
accurate checks regarding the solar system tests should still be
carried out. Nevertheless, even if some problems could be
encountered, there will always remain a reasonable chance that the
situation improves by taking into account higher-order string
corrections, as has been wisely indicated in Ref.~\cite{sami}. The
introduction of such higher-order string corrections in the above
scenario will be discussed elsewhere.

\section*{Acknowledgements}

Thanks are given to M.~Sami and to M.~Sasaki for very interesting
discussions. This paper is an outcome of the collaboration program
INFN (Italy)--DGICYT (Spain). It has been also supported in part by
MEC (Spain), projects BFM2003-00620 and PR2006-0145, by JSPS (Japan)
XXI century COE program of Nagoya University project 15COEEG01, by
Monbusho grant no.18549001 (Japan), by LRSS project N4489.2006.02,
by RFBR grant 06-01-00609 (Russia), and by AGAUR (Gene\-ra\-litat de
Catalunya), contract 2005SGR-00790.

\appendix

\newcommand{\s}[1]{\section{#1}}
\renewcommand{\ss}[1]{\subsection{#1}}
\newcommand{\sss}[1]{\subsubsection{#1}}
\def\M{{\cal M}} %%% calligraphic M
\newcommand{\ca}[1]{{\cal #1}} %%% calligraphic
\def\segue{\qquad\Longrightarrow\qquad} %%% it follows
\def\prece{\qquad\Longleftarrow\qquad} %%% it follows
\def\hs{\qquad} %%% horizontal space
\def\nn{\nonumber} %%% no number for eqnarray
\def\beq{\begin{eqnarray}} %%% begequation/eqnarray
\def\eeq{\end{eqnarray}} %%% endequation/eqnarray
\def\ap{\left.} %%% open bracket
\def\at{\left(} %%% open (
\def\aq{\left[} %%% open [
\def\ag{\left\{} %%% open {
\def\cp{\right.} %%% close bracket
\def\ct{\right)} %%% close )
\def\cq{\right]} %%% close ]
\def\cg{\right\}} %%% close }

\def\R{{\hbox{{\rm I}\kern-.2em\hbox{\rm R}}}} %%% real numbers
\def\H{{\hbox{{\rm I}\kern-.2em\hbox{\rm H}}}} %%% Hilbert space
\def\N{{\hbox{{\rm I}\kern-.2em\hbox{\rm N}}}} %%% natural numbers
\def\C{{\ \hbox{{\rm I}\kern-.6em\hbox{\bf C}}}} %%% complex numbers
\def\Z{{\hbox{{\rm Z}\kern-.4em\hbox{\rm Z}}}} %%% integers numbers
\def\ii{\infty} %%% infinit
\def\X{\times\,} %%% times
\newcommand{\bin}[2]{\left(\begin{matrix}{#1\cr #2\cr}
\end{matrix}\right)} %%% binomial
\newcommand{\fr}[2]{\mbox{$\frac{#1}{#2}$}} %%% small fraction
\def\Det{\mathop{\rm Det}\nolimits} %%% Determinant
\def\tr{\mathop{\rm tr}\nolimits} %%% trace
\def\Tr{\mathop{\rm Tr}\nolimits} %%% Trace
\def\rot{\mathop{\rm rot}\nolimits} %%% Rotore
\def\PP{\mathop{\rm PP}\nolimits} %%% Finite part
\def\Res{\mathop{\rm Res}\nolimits} %%% Residue
\def\res{\mathop{\rm res}\nolimits} %%% Residue
\renewcommand{\Re}{\mathop{\rm Re}\nolimits} %%% Real
\renewcommand{\Im}{\mathop{\rm Im}\nolimits} %%% Imaginary
\def\dir{/\kern-.7em D\,} %%% Dirac Operator
\def\lap{\Delta\,} %%% Laplacian
\def\arccosh{\mbox{arccosh}\:} %%% hyperbolic
\def\arcsinh{\mbox{arcsinh}\:} %%% functions
\def\arctanh{\mboz{arctanh}\:} %%%
\def\arccoth{\mbox{arccoth}\:} %%%
%%%%% GREEK ALPHABET
\def\al{\alpha}
\def\ga{\gamma}
\def\de{\delta}
\def\ep{\varepsilon}
\def\ze{\zeta}
\def\io{\iota}
\def\ka{\kappa}
\def\la{\lambda}
\def\ro{\varrho}
\def\si{\sigma}
\def\om{\omega}
\def\ph{\varphi}
\def\th{\theta}
\def\te{\vartheta}
\def\up{\upsilon}
\def\Ga{\Gamma}
\def\De{\Delta}
\def\La{\Lambda}
\def\Si{\Sigma}
\def\Om{\Omega}
\def\Te{\Theta}
\def\Th{\Theta}
\def\Up{\Upsilon}

\def\h{\hat h}

\section{One-loop effective action in $F(G)$ gravity on
de Sitter space}

The important issue in any fundamental gravitational theory is stability
issue, as it indicates if some of the highly symmetric spaces (flat, de
Sitter or Anti-de Sitter) could be the ground state.
Having in mind that de Sitter space appears as classical solution of the
above theory in the early as well as in the late universe we study
stability of GB modified gravity in de Sitter space.

In order to study the
stability of $F(G)$ gravity, one possibility
is to calculate the one-loop effective action on the corresponding
background. It will be here shown that, in the calculation of the
one-loop effective action in higher-derivative gravity a multiplicative
anomaly appears naturally. Thus, before starting the calculation we
first review this concept in a general situation.
The usual definition of the determinant of an operator,
$A$, is done through its zeta function \cite{zb1} $\zeta_A (s) =
\sum_{i\in I} \lambda_i^{-s} = \tr \, A^{-s}$,
where analytic continuation in $s$ to cover the domain of the
complex plain to the left of the abscissa of convergence $s=s_0$
is naturally assumed. Recall that $s_0=m/n$, namely it
is given by the dimension of
the working space, $m =$ dim $M$, divided by the order of the operator,
$n =$ ord $A$.
For an elliptic operator of positive order on a compact manifold, satisfying the
usual spectral condition (e.g. the presence of an
Agmon-Nirenberg cut in the spectrum), the zeta
function exists and has very nice general properties, as being
meromorphic with possible single poles on integer and
specific fractional positions
$
s_k = (n-k)/m, \ k=0,1,2,\ldots,n-1,n+1, \dots
$
of the negative real axis only \cite{ze1a}.
The determinant of $A$ is then given by \cite{rs1}
\beq {\det}_\zeta A = \exp \left[ -\zeta_A ' (0)
\right]. \eeq
This definition only depends on the homotopy class of the spectral cut.

Now, given the operators $A$, $B$ and $AB$, even if
$\zeta_A$, $\zeta_B$ and $\zeta_{AB}$ exist, it turns out that, in
general,
$
{\det}_\zeta (AB) \neq {\det}_\zeta A \ {\det}_\zeta B.
$
The multiplicative, or noncommutative, or determinant anomaly
(also called defect of the determinant) is defined as:
\beq
\delta (A,B) = \ln \left[ \frac{
\det_\zeta (AB)}{\det_\zeta A \ \det_\zeta B} \right] =
-\zeta_{AB}'(0)+\zeta_A'(0)+\zeta_B'(0). \label{an1}
\eeq
There is a useful formula due to Wodzicki for the multiplicative
anomaly \cite{wodz87b,kass1}. In terms of Wodzicki's residue:
res $A=2$ Res$_{s=0}\ \tr (A \Delta^{-s})$, $\Delta$ being the
Laplacian operator, or equivalently, in local form using the
symbol expansion,
$\mbox{res} \ A = \int_{S^*M} \mbox{tr}\ a_{-n}(x,\xi)
\, d\xi $, with $S^*M \subset T^*M $ the co-sphere bundle on $M$,
the multiplicative anomaly can be obtained as
\beq
\delta (A,B)= \frac{\mbox{res}\left\{ \left[ \ln \sigma (A,B)\right]^2
\right\}}{2 \mbox{ ord}\, A \mbox{ ord}\, B
(\mbox{ord}\, A + \mbox{ord}\, B)}, \qquad \quad
\sigma (A,B) := A^{\mbox{ord}\, B} B^{-\mbox{ord}\, A}.
\eeq
Several implications of this multiplicative anomaly have been
extensively discussed in the literature \cite{evz12}. Here we
consider a very different situation, never met before, namely
the case when
one of the two operators slowly vanishes owing to the change of a
certain parameter, a situation that may be not too unfrequent.
The main issue here is to check if, in fact, the shrinking operator
could leave some imprint on the surviving operator, through the
multiplicative anomaly term \cite{evz12}. To simplify notation
we will drop the $\zeta$ from the determinants hereafter.

Such anomaly appears basically in higher derivative gravities, like
the $R^2$ theory, when one calculates the one-loop effective action
in some background. This will precisely be case under consideration
here. The anomaly may persist even after one of the operators shrinks,
say as a consequence of time evolution or by the action of some
other parameter, as we are now going
to see. In fact, as an example, let us consider the uniparametric
family of operators $B_\epsilon$, where $\epsilon$ is a real
parameter that will eventually go to zero, e.g., adiabatically. For
further simplicity, let us just take that $B_\epsilon = \epsilon B$,
an overall constant factor. It is easy to see from its definition that
the anomaly $\delta (A,B_\epsilon)$ does not actually depend on
$\epsilon$, in fact
\beq \delta (A,B_\epsilon) = \delta (A,B). \eeq
And, again from the definition of the anomaly (\ref{an1}), it
turns out that in the limit when the $B_\epsilon$ operator
adiabatically disappears, a contribution may remain, in the way that
\beq
\left. \frac{\det (A B_\epsilon)}{\det B_\epsilon}\right|_{\epsilon
\to 0} = \det A \cdot e^{ \delta (A,B)}.
\eeq
This is again an implication of the presence of the
multiplicative anomaly and, on its turn, of the definition of the
zeta determinant. It contributes, generically, an additional term
to the determinant of the resulting operator after taking the limit.

Now to the physics. Corresponding to the scalar version of $F(G)$
gravity, we have the Euclidean action
\beq
S_E[\tilde g]=-\frac{1}{\ka^2}\int\:d^4x\,\sqrt{\tilde g}\aq
\,\tilde R-\frac{\ep\,\tilde g^{ij}\partial_i\tilde\phi\partial_j\tilde\phi}2
+F(\tilde\phi)+F'(\tilde\phi)(\tilde G-\tilde\phi)\cq\,,
\label{Action}
\eeq
which is equivalent to the modified Gauss-Bonnet gravity as
investigated in \cite{Cognola:2006eg}.
In (\ref{Action}) $\tilde\phi$ is a scalar field which ``on shell''
becomes equal to the Gauss-Bonnet invariant $\tilde G$,
$\tilde g_{ij}$ is the metric, $\tilde R$ the related scalar
curvature
and finally $F(\tilde\phi)$ is an arbitrary smooth potential.
An interesting observation is that the multiplicative anomaly
commutes with the limit of the small parameter $\epsilon \to 0$.
We will confirm that this is the case below.
It may persist after the limit is enforced.

As it as been shown in \cite{Cognola:2006eg},
to ensure the existence of de Sitter solutions
the function $F$ has to satisfy the condition
\beq
GF'(G)-F(G)=\frac{R}{2}\,,
\label{dSC}
\eeq
where the quantities $G$ and $R$ are respectively
the Gauss-Bonnet invariant and
the scalar curvature related to the de Sitter constant solution.

Now we are going to consider small fluctuations around the constant
curvature solution $g_{ij}$, associated with de Sitter solution, so
we put
\beq
\tilde g_{ij}=g_{ij}+h_{ij}\:,\hs\hs h=g^{ij}h_{ij}\:,
\hs\hs\tilde\phi=G+\phi\,, \eeq and we develop the action
(\ref{Action}) around the constant curvature solution up to second
order in $h_{ij}$ and $\phi$. As usual all tensor indices are
lowered and risen by means of $g_{ij}$. After a straightforward
calculation, taking into account that we are dealing with a
maximally symmetric space, for the quadratic part and disregarding
total derivatives we obtain \beq {\cal L}_2&=&
{\frac{-{}{f_{0}}\,{{\h_{ij}}^2}{}}{4}}-{\frac{R\,{{\h_{ij}}^2}}{6}}
+{\frac{{f_{1}}\,{R^2}\,{{\h_{ij}}^2}}{6}}+{\frac{\h_{ij}\,
\Delta\,\h_{ij} }{4}} +{\frac{{f_{1}}\,R\,\h_{ij}\, \Delta\,\h_{ij}
}{6}} \nonumber\\ &&
+{\frac{{f_{0}}\,R\,{{\xi_{i}}^2}}{8}}+{\frac{{R^2}\,{{\xi_{i}}^2}}{16}}
-{\frac{31\,{f_{1}}\,{R^3}\,{{\xi_{i}}^2}}{288}}+{\frac{{f_{0}}\,\xi_{i}\,
\Delta\,\xi_{i} }{2}} \nonumber\\ && +{\frac{R\,\xi_{i}\,
\Delta\,\xi_{i} }{4}}-{\frac{17\,{f_{1}}\,{R^2}\,\xi_{i}\,
\Delta\,\xi_{i} }{36}}-{\frac{{f_{1}}\,R\,\xi_{i}\, \Delta\,
\Delta\,\xi_{i} }{6}} \nonumber\\ &&
+{\frac{{f_{0}}\,{h^2}}{16}}-{\frac{{f_{2}}\,{{\phi}^2}}{2}}+{\frac{{f_{1}}\,{h^2}\,{R^2}}{48}}
-{\frac{{f_{2}}\,h\,\phi\,{R^2}}{6}}-{\frac{3\,h\,
\Delta\,h }{32}} \nonumber\\ && +{\frac{5\,{f_{1}}\,h\,R\,
\Delta\,h }{32}}+{\frac{\epsilon\,\phi\, \Delta\,\phi
}{2}}-{\frac{{f_{2}}\,h\,R\, \Delta\,\phi }{4}} \nonumber\\ &&
+{\frac{h\,R\, \Delta\,\sigma }{16}}-{\frac{{f_{1}}\,h\,{R^2}\,
\Delta\,\sigma }{16}}+{\frac{{f_{2}}\,\phi\,{R^2}\, \Delta\,\sigma
}{12}} \nonumber\\ && -{\frac{{f_{0}}\,R\,\sigma\, \Delta\,\sigma
}{16}}-{\frac{{R^2}\,\sigma\, \Delta\,\sigma
}{32}}+{\frac{17\,{f_{1}}\,{R^3}\,\sigma\, \Delta\,\sigma }{288}}
\nonumber\\ && +{\frac{3\,h\, \Delta\, \Delta\,\sigma
}{16}}-{\frac{3\,{f_{1}}\,h\,R\, \Delta\, \Delta\,\sigma
}{16}}+{\frac{{f_{2}}\,\phi\,R\, \Delta\, \Delta\,\sigma }{4}}
\nonumber\\ && -{\frac{3\,{f_{0}}\,\sigma\, \Delta\, \Delta\,\sigma
}{16}}-{\frac{R\,\sigma\, \Delta\, \Delta\,\sigma
}{8}}+{\frac{3\,{f_{1}}\,{R^2}\,\sigma\, \Delta\, \Delta\,\sigma
}{16}} \nonumber\\ && -{\frac{3\,\sigma\, \Delta\, \Delta\,
\Delta\,\sigma }{32}}+{\frac{{f_{1}}\,R\,\sigma\, \Delta\, \Delta\,
\Delta\,\sigma }{32}}\,, \label{L2}\eeq where $f_k=F^{(k)}(G)$
($k=0,1,2$) is the derivative of the function $F(\tilde\phi)$
evaluated at $G$, while $\h_{ij}$, $\xi_k$, $\sigma$ and $h$ are the
irreducible components of the symmetric tensor field $h_{ij}$. They
are related by \beq
h_{ij}=\h_{ij}+\nabla_i\xi_j+\nabla_j\xi_i+\nabla_i\nabla_j\sigma
+\frac14\,g_{ij}(h-\lap\sigma)\:, \label{tt} \eeq and satisfy the
conditions \beq \nabla_k\xi^k=0\:,\hs\hs \nabla_i\h^{ij}=0\:,\hs\hs
g^{ij}\h_{ii}=0\:. \label{AAA4} \eeq Here $\nabla_k$ and
$\lap=g^{ij}\nabla_i\nabla_j$ are respectively the covariant
derivative and the Laplace-Beltrami operators related to the de
Sitter metric $g_{ij}$. It has to be remarked that Eq.~(\ref{L2})
coincide with the analog expression written in \cite{Cognola:2006eg}
when one puts $\phi=\tilde G$ and $\epsilon=0$ (see Sec. IV in the
cited paper).

As it is well known, invariance under diffeormorphisms renders the
operator in the $(h,\si)$ sector not invertible. One needs a gauge
fixing term and a corresponding ghost compensating term.
We consider the class of gauge conditions
\beq
\chi_k=\nabla_j h_{jk}-\frac{1+\rho}4\,\nabla_k\,h\:,
\nn
\eeq
parametrized by the real parameter $\rho$.
As gauge fixing we choose the standard expression
\beq
{\cal L}_{gf}=\frac12\,\chi^iG_{ij}\chi^j\,,\hs\hs
G_{ij}=\al g_{ij}
\label{AAA5}
\eeq
and so the corresponding ghost Lagrangian reads \cite{buch}
\beq
{\cal L}_{gh}=G_{ij}B_i\frac{\de\,\chi^j}{\de\,\ep^k}C^k\,,
\label{AAA6}
\eeq
where $C_k$ and $B_k$ are the ghost and anti-ghost
vector fields, respectively, while $\de\,\chi^k$ is the variation of
the gauge condition due to an infinitesimal gauge transformation of
the field. It reads
\beq
\de\,h_{ij}=\nabla_i\ep_j+\nabla_j\ep_i\segue
\frac{\de\,\chi^i}{\de\,\ep^j}=g_{ij}\,\lap+R_{ij}+\frac{1-\rho}{2}\,\nabla_i\nabla_j\,.
\label{AAA7}
\eeq
Neglecting total derivatives one has
\beq
{\cal L}_{gh}=\al\,B^k\,\at\lap+\frac{R}{4}\ct\,C_k
\label{AAA8}
\eeq
and finally in irreducible components we obtain
\begin{eqnarray}
{\cal L}_{gf} &=&\frac{\al}2\aq\xi^k\,\at\lap+\frac{R}4\ct^2\,\xi_k
+\frac{3\rho}{8}\,h\,\at\lap+\frac{R}3\ct\,\lap\,\si
\cp
\nn\\&&\hs\ap
-\frac{\rho^2}{16}\,h\,\lap\,h
-\frac{9}{16}\,\si\,\at\lap+\frac{R}3\ct^2\,\lap\,\si\cq\,,
\label{AAA10}
\eeq
\beq
{\cal L}_{gh} &=&\alpha\aq\hat B^i\at\lap+\frac{R}{4}\ct\hat C^j
+\frac{\rho-3}{2}\,b\,\at\lap-\frac{R}{\rho-3}\ct\,\lap c\cq\,,
\eeq
where ghost irreducible components are defined by
\beq
C_k&=&\hat C_k+\nabla_k c\,,\hs\hs \nabla_k\hat C^k=0\,,
\nn\\
B_k&=&\hat B_k+\nabla_k b\,,\hs\hs \nabla_k\hat B^k=0\,.
\label{AAA11}
\eeq
In order to compute the one-loop contributions to
the effective action one has to consider the path integral for the
bilinear part
\beq
{\cal L}= {\cal L}_2+\,{\cal L}_{gf}+{\cal L}_{gh} \label{AAA12}
\eeq
of the total Lagrangian and take into
account the Jacobian due to the change of variables with respect to
the original ones. In this way, one gets \cite{frad,buch}
\beq
Z^{(1)}&=&\at\det G_{ij}\ct^{-1/2}\,\int\,D[h_{ij}]D[C_k]D[B^k]\:
\exp\,\at -\int\,d^4x\,\sqrt{g}\,{\cal L}\ct
\nn\\
&=&\at\det G_{ij}\ct^{-1/2}\,\det J_1^{-1}\,\det J_2^{1/2}\,
\nn\\
&&\times \int\,D[h]D[\h_{ij}]D[\xi^j]D[\si] D[\hat
C_k]D[\hat B^k]D[c]D[b]\:\exp\, \at-\int\,d^4x\,\sqrt{g}\,{\cal
L}\ct\,,
\eeq
where the determinant of the operator $G_{ij}$ for our choice
is trivial, while
$J_1$ and $J_2$ are the Jacobians due to the
change of variables in the ghost and tensor sectors respectively.
They read \cite{buch}
\beq
J_1=\lap_0\,,\hs\hs
J_2=\at-\lap_1-\frac{R}{4}\ct\at-\lap_0-\frac{R}{3}\ct\,\lap_0\,,
\label{AAA13}
\eeq
$\lap_0$ and $\lap_1$ being the Laplacians acting on scalar and vector
fields.

Now, a straightforward computation (disregarding the multiplicative
anomaly (\cite{Elizalde:1998vd})) leads to the following off-shell
one-loop contribution to the ``partition function''
\beq
e^{-\Ga^{(1)}}&\equiv&Z^{(1)}=\det\at-\lap_1-\frac{R}{4}\ct^{1/2}
\:\det\at-\lap_0-\frac{R}{2}\ct
\nn\\ &&\hs\hs\times\,
\det\aq-\lap_2-\frac{R(9f_0+4R-12X)}{3(4f_0+3R-4X)}\cq^{-1/2}
\nn\\ &&\hs\hs\hs\times\,
\det\aq
\at{-\frac{9\,{\epsilon}}{256}}-{\frac{21\,{\epsilon}\,{f_{0}}}{256\,R}}
+{\frac{{{{f_{2}}}^2}\,{R^2}}{512}}
+{\frac{21\,{\epsilon}\,X}{256\,R}}\ct\,\lap_0^3
\cp\nn\\ &&+\at
{-\frac{17\,{\epsilon}\,{f_{0}}}{256}}+{\frac{9\,{f_{2}}}{256}}+{\frac{21\,{f_{0}}\,{f_{2}}}{256\,R}}
-{\frac{7\,{\epsilon}\,R}{256}}+{\frac{7\,{{{f_{2}}}^2}\,{R^3}}{1536}}+{\frac{31\,{\epsilon}\,X}{512}}
-{\frac{21\,{f_{2}}\,X}{256\,R}}\ct\,\lap_0^2
\nn\\&&+\at
{\frac{17\,{f_{0}}\,{f_{2}}}{256}}-{\frac{9\,{\epsilon}\,{f_{0}}\,R}{1024}}+{\frac{7\,{f_{2}}\,R}{256}}
-{\frac{3\,{\epsilon}\,{R^2}}{1024}}+{\frac{5\,{{{f_{2}}}^2}\,{R^4}}{1536}}-{\frac{31\,{f_{2}}\,X}{512}}
+{\frac{3\,{\epsilon}\,R\,X}{512}}\ct\,\lap_0
\nn\\&&\hs\hs +\at
{\frac{9\,{f_{0}}\,{f_{2}}\,R}{1024}}+{\frac{3\,{f_{2}}\,{R^2}}{1024}}+{\frac{{{{f_{2}}}^2}\,{R^5}}{1536}}
\ap-{\frac{3\,{f_{2}}\,R\,X}{512}}\ct\,
\cq^{-1/2}\,.
\label{PF}
\eeq
For convenience we have written this in the Landau gauge corresponding to
$\rho=1$ and $\al=\ii$ and moreover we have put
$X=f_0+R/2-Gf_1$. Thus, $X=0$ is the ``on-shell'' condition
corresponding to the de Sitter solution.

As a check one can easily see that Eq.~(\ref{PF}) has the correct
limit in the Einstein plus cosmological constant case, that is
when $\ep=0,\,f_2=0,\,f_0=-1/2$. In fact in such a case
the complicated expression in the scalar sector decouples and one obtains
the well known result reported in Ref.~ \cite{frad}.
Furthermore, when $\ep=0$, it can be shown that also at one-loop level
such a model is equivalent to
the $F(G)$ theory developed in \cite{Cognola:2006eg}, which confirms
the commutativity with the multiplicative anomaly.

\section{Stability criteria for modified Gauss-Bonnet
gravity}

In this Appendix, applying the results of the previous Appendix we
study the stability issue for several specific models.
In principle, one can study the one-loop effective action for
the de Sitter space explicitly in terms of special functions
\cite{frad}. This happens in some cases, where one is able to express
the determinant, say det $(-\Delta_0^3+ a_1 \Delta_0^2-a_2
\Delta_0+a_3)$, in terms of a product of more elementary determinants
of lower-dimensional
operators. It is there that the multiplicative anomaly explicitly
appears. Fortunately, for the case under discussion we are able to
carry out a detailed analysis of the stability conditions of the
model without the need to go through all this cumbersome process.
This is what we do here.

In reference \cite{Cognola:2005sg}, the one-loop effective action
has been also used in deriving a stability criterion for the class
of modified models described by the function $f(R)$.
Such a criterion of stability has been obtained by imposing
the absence of negative eingenvalues related to the
Laplace like operators appearing in the
regularized one-loop effective action and it has
been independently confirmed in \cite{faraoni}
within a pure classical approach, namely, involving the study of
cosmological perturbations.

In the case we are discussing here the situation is more complicated
from the technical view point, since the functional determinant one
has to investigate involves an algebraic polynomial of third order
in the Laplace operator. But, in principle, a stability criterion
for such a class of modified Gauss-Bonnet models could be
investigated along the same lines, namely by requiring the
vanishing of the imaginary part in the one-loop effective action or
the absence of negative eingenvalues. Note that the equivalent
analysis of cosmological perturbations in $F(G)$ gravity is
extremely complicated\cite{Mota}. However, such study may be very
important cosmologically, since it may prove (or disprove) the usual
belief that the current (almost) de Sitter dark energy era may be
eternal.

In order to discuss this issue we take the limit $\ep\to0$
and consider the on-shell condition $X=0$. In principle
the operator of the scalar sector in (\ref{PF}) may be written
in the factorized form
\beq
A(X)=X^3+a_1X^2+a_2X+a_3=(X-X_1)(X-X_2)(X-X_3)\,,
\eeq
where now $X=-\lap_0$ is a non-negative differential operator,
$X_i$ are the roots of the third order algebraic equation
$A(X)=0$ and the constant coefficients $a_i$ are given by
\beq
a_1&=&-R\at\frac{7}{3}+\frac{18}{f_2R^3}+\frac{42f_0}{f_2R^4}\ct\,,
\nn\\
a_2&=&R^2\at\frac53+\frac{14}{f_2R^3}+\frac{34f_0}{f_2R^4}\ct\,,
\\
a_3&=&-R^3\at\frac13+\frac{3}{2f_2R^3}+\frac{9f_0}{f_2R^4}\ct\,.
\nn\eeq
The nature of the roots depend on the
discriminant, $D$, which reads
\beq
D=\at\frac{3a_2-a_1^2}{9} \ct^3+\at \frac{9a_1a_2-27a_3-2a_1^3}{54}\ct ^2\,.
\eeq
Depending on the sign of the discriminant one has the three cases:
\begin{description}
\item{(i)} $D>0$, then one root is real while the other two form a couple
of conjugated complex numbers, say $X_1=\bar X_3$.
\item{(ii)} $D=0$, then all the roots are real and at least two
of them are equal, say $X_2=X_3$.
\item{(iii)} $D<0$, then the three roots are real and distinct.
\end{description}
As a consequence, in the first two cases the factorization reads
$A(X)=(X-X_1)B(X)$,
$B(X)$ being a non negative differential operator and
then one has stability of the Gauss-Bonnet modified
models provided $X_1<0$ or, which is equivalent, $a_3>0$, since
$X_1X_2X_3=-a_3$.
Then requiring $a_3>0$ and $D\geq0$ we arrive at the following
sufficient conditions for the stability of the
model we are considering on the de Sitter background:
\beq
&&6f_0-R^2f_1-3R=0\,,\nn\\
&&\frac{27f_0}{f_2}+\frac{9R}{f_2}+2R^4<0\,,\nn\\
&&{{76204800\,{{{f_0}}^4}}}+{{136950912\,{{{f_0}}^3}\,R}}+{{92161152\,{{{f_0}}^2}\,{R^2}}}\nonumber\\
&&\hs+{{27527040\,{f_0}\,{R^3}}}+{{3079296\,{R^4}}}+{{6901200\,{{{f_0}}^3}\,{f_2}\,{R^4}}}\nonumber\\
&&\hs\hs+{{9430128\,{{{f_0}}^2}\,{f_2}\,{R^5}}}+{{4292784\,{f_0}\,{f_2}\,{R^6}}}\nonumber\\
&&\hs\hs\hs+{{651024\,{f_2}\,{R^7}}}+{{171975\,{{{f_0}}^2}\,{{{f_2}}^2}\,{R^8}}}+{{154794\,{f_0}\,{{{f_2}}^2}\,{R^9}}}\nonumber\\
&&\hs\hs\hs\hs+{{34815\,{{{f_2}}^2}\,{R^{10}}}}+{{1600\,{f_0}\,{{{f_2}}^3}\,{R^{12}}}}+{{704\,{{{f_2}}^3}\,{R^{13}}}}
\leq 0\,. \label{S}\eeq The first one is the on-shell condition,
while the second and the third ones derive from $a_3>0$ and $D\geq
0$ respectively.

When the discriminant is negative all the roots are distinct and
the condition $a_3>0$ does not ensure the operator $A(X)$ to be
non-negative. Of course, a sufficient condition can be obtained
by requiring all the roots to be negative, but in such a case,
for technical reasons we did not find
a reasonable stable model.

As an example let us consider the choice
$F(\tilde G)=\al\sqrt{\tilde G}+\beta\tilde G^\ga$,
$\al$ and $\ga$ being arbitrary dimensionless parameters,
while $\beta$ has dimensions $[\mbox{mass}]^{1-2\ga}$.
This is the only dimensional parameter in the model and for this reason
it remains free also after imposing the above restrictions.
The on shell condition gives
\beq
R^{2\ga-1}=\frac{6^{\ga-1}(6+\sqrt6\,\al)}{2\beta}\,,
\eeq
where $\al,\beta,\ga$ are assumed to have
values such that $R>0$, since we want a de Sitter solution.
The other two conditions above fix the possible ranges of
$\al$ and $\ga$. In particular it can be seen that there
are stable solutions for very small,
positive values of both $\al$ and $\ga$, only.

A careful analysis of the stability conditions (\ref{S}) leads to
the following conclusions. First, the whole analysis can be done
in terms of $f_0$ and $f_2$, by the substitution of the first equation,
namely
\beq f_1= \frac{3}{R} \left(\frac{2 f_0}{R}-1 \right).\eeq
Then, an
analysis of the roots of the last inequality, seen as
a polynomial of $f_0$ at the bound of the second inequality, shows
that there is a stability region for values of $f_0$ around
$f_0\simeq -R/2$, that is $f_2 \simeq -6/R$, and $f_2 \leq 9/4R^3$.
This is depicted in the figures. All of them are plots of the last
inequality in (\ref{S}), where the relevant domain for $f_0$ is
identified (as we see there, an additional one exists but it requires
very precise fine tuning). Fig.~1a
\begin{figure}[th]
%\vskip-3cm
\centerline{\epsfxsize=9cm \epsfbox{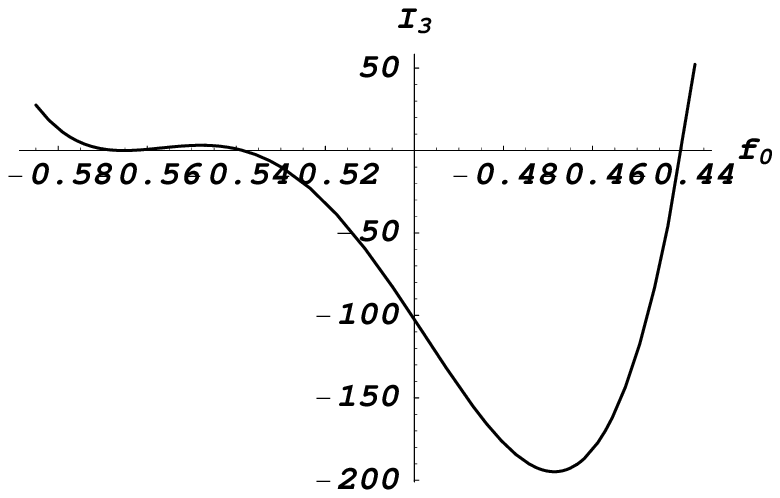} \ \ \
\epsfxsize=9cm \epsfbox{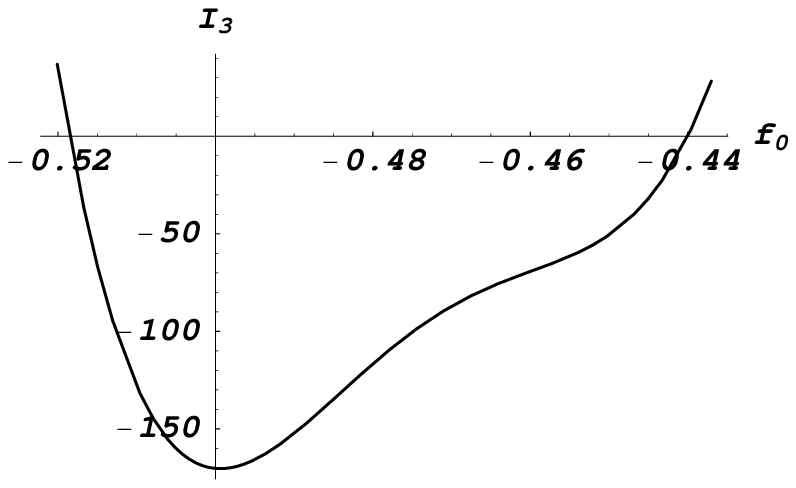}}
%\vskip-25mm
\caption{{\protect\small
Fig.~1a, left, plot of the lhs of the last
inequality in (\ref{S}), which we call $I_3$, where $f_2$ lies
on the left border of
the first stability domain (\ref{fineq1}). Fig.~1b, right, same plot
where $f_2= 1/R^3$ lies inside the region delimited by the
first of the inequalities in (\ref{S}), not far from the right end
of the first stability domain}}
\label{f1}
\end{figure}
is obtained for $f_2$ on the first border
of the first of the inequalities in (\ref{S}), which is two-sided,
namely
\beq f_2 <(-9/2R^3)(1+3f_0/R), && f_0<-R/3, \label{fineq1} \\
f_2 >(-9/2R^3)(1+3f_0/R), && f_0>-R/3.\label{fineq2}\eeq
Fig.~1b shows the same plot but for $f_2$ well inside the region
delimited by the
first of the inequalities: $f_2= 1/R^3$. Fig.~2a shows the
stability region corresponding to the very particular case of
a vanishing second derivative of $F(G)$, namely
$f_2=0$, while Fig.~2b corresponds to the value $f_2= -1/R^3$,
which marks quite closely the end of the first stability domain
(\ref{fineq1}) for $f_2$. We thus see that for very reasonable values
of the first derivatives of the function $F(G)$, we are inside the
domain of convergence. To identify specific functions having these
derivatives, however, is not an easy task, as the preceding example
shows.
\begin{figure}[htb]
%\vskip-3cm
\centerline{\epsfxsize=9cm \epsfbox{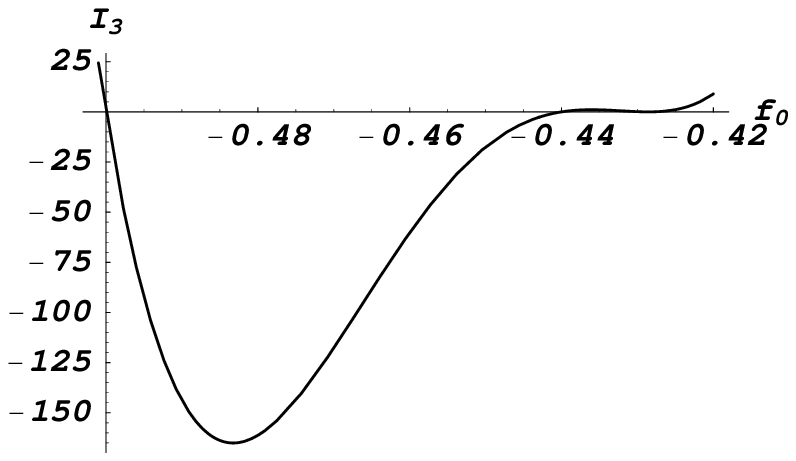} \ \ \
\epsfxsize=9cm \epsfbox{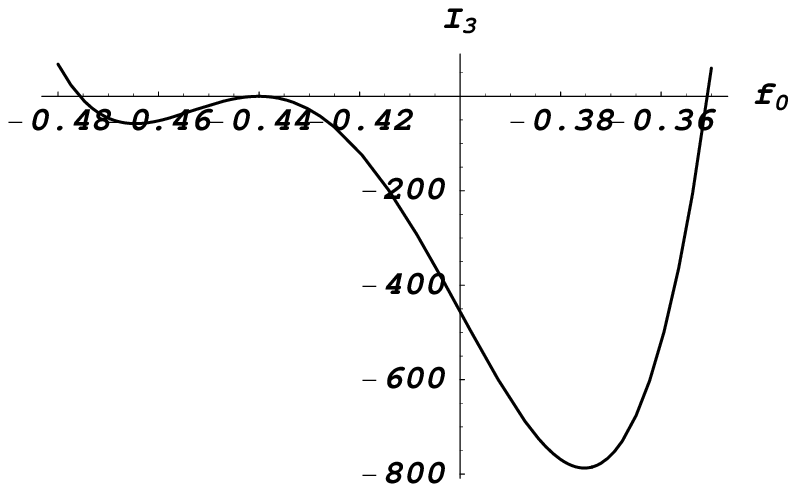}}
%\vskip-25mm
\caption{{\protect\small
Fig.~2a, left, plot of $I_3$, where $f_2$ corresponds
to the case $f_2=0$. Fig.~2b, right, same plot
where $f_2=-1/R^3$ lies close to the left end of the second
stability domain (\ref{fineq2}).}}
\label{f2}
\end{figure}

Similarly, one can study the stability conditions of the
de Sitter universe which corresponds to the final era in the first
of the reconstruction scenarios considered above,
for other versions of string-inspired gravity. It is our impression,
that this study, which recently became quite important due to
the possible emergence of a (metastable) de Sitter vacuum in string
theory, could prove to be less involved than
the corresponding analysis of cosmological perturbations \cite{Mota}.
Indeed, even the classical stability study of higher-derivative
gravity on the de Sitter background is already not that easy
\cite{lesha}.

\end{document}